\documentclass[pra,floatfix,twocolumn,nobalancelastpage]{revtex4-1}

\usepackage{graphicx}
\usepackage{braket}
\usepackage{amsmath}
\usepackage{amsthm,amsmath,amsfonts,dsfont}
\usepackage{diagbox}
\usepackage{cancel}
\usepackage{bm}

\usepackage{color}

\newcommand{\comment}[1]{}

\newcommand{\eref}{Eq.~\eqref}
\newcommand{\tr}[2][]{\text{Tr}_{#1}\left\{#2\right\}}

\begin{document}
	
	\title{Variational analysis of driven-dissipative bosonic fields}
	
	\author{Tim Pistorius}
	\affiliation{Institut f\"ur Theoretische Physik, Leibniz Universit\"at Hannover, Appelstra{\ss}e 2, 30167 Hannover, Germany}
	\author{Hendrik Weimer}
	\email{hweimer@itp.uni-hannover.de}
	\affiliation{Institut f\"ur Theoretische Physik, Leibniz Universit\"at Hannover, Appelstra{\ss}e 2, 30167 Hannover, Germany}

	\begin{abstract}
          We present a method to perform a variational analysis of the quantum master equation for driven-disspative bosonic fields with arbitrary large occupation numbers. Our approach combines the P representation of the density matrix and the variational principle for open quantum system. We benchmark the method by comparing it to wave-function Monte-Carlo simulations and the solution of the Maxwell-Bloch equation for the Jaynes-Cummings model. Furthermore, we study a model describing Rydberg polaritons in a cavity field and introduce an additional set of variational paramaters to describe correlations between different modes.
          
	\end{abstract}
	
	
	\maketitle
	
	
	\section{Introduction}

The theoretical analysis of driven-dissipative quantum many-body
systems is a very challenging task, as many methods developed for
closed quantum systems cannot be applied. While important insights
have been obtained using the Wave-Function Monte-Carlo (WFMC) method
\cite{Dalibard1992,Dum1992} or tensor network approaches \cite{Cui2015,Mascarenhas2015,Mendoza-Arenas2016,Kshetrimayum2017}, the numerical
study of bosonic fields with large occupation numbers remains an
outstanding problem \cite{Weimer2020}. Here, we show that the
variational principle for open quantum systems \cite{Weimer2015}
provides for a natural way to represent arbitrarily large occupation
numbers in terms of a small set of variational parameters, thus
providing a highly efficient description of the system.

Driven-dissipative systems of bosons arise in many different settings,
ranging from semiconductor polaritonic systems \cite{Carusotto2013},
to cavity quantum electrodynamics arrays \cite{Aron2016}, to Rydberg
polaritons in atomic quantum gases
\cite{Dudin2012,Peyronel2012}. Especially for the latter, the
development of efficient numerical descriptions is of great interest
due to the applications of Rydberg polaritons in the context of
strongly correlated photon states \cite{Jia2018} and photonic quantum
computing \cite{Gorshkov2011a,Tiarks2019,Bienias2020}.

In this article, we present a variational treatment of the steady
state of driven-dissipative bosons. While the Lindblad formalism
allows to calculate the steady state based on the solution of a
quantum master equation \cite{Breuer2002}, a brute-force solution
becomes prohibitive for large Hilbert space dimensions. Here, we apply
the variational principle for open quantum systems, which already
proved to be a reliable method in the context of spin-$\frac{1}{2}$
particles \cite{Weimer2015,Weimer2015a,Overbeck2016}. However, a
direct implementation of the variational principle for bosonic fields
results in a large amount of variational parameters to describe large
Hilbert space, which drastically reduces the efficiency of the method
and still relies on a cut-off of the Hilbert space. Therefore, we turn
to a different implementation of the variational principle based on
the Glauber-Sudarshan P representation.
        
Phase-space representations of the density matrix have received
considarble interest in recent years to classify the nonclassical
properties of quantum states
\cite{Ryl2015,Bohmann2020,Drummond1980,Kim2002,Zavatta2007,Lvovsky2002,Agarwal1992}. For
this, the P representation by Glauber and Sudarshan
\cite{Glauber1963,Sudarshan1963} is particularly useful, as negative
values directly point to nonclassical behavior.

In this following, we want to use this representation of the density
matrix to expand the variational method for open quantum systems to
bosonic fields. We also show how the formalism of the variational
principle translates to the Heisenberg equations of motions and how
this can be used to extend the method to highly singular P
distribution where an explicit representation is not
feasible. We benchmark our approach by comparing to WFMC simulations \cite{Johansson2012,Johansson2013} mean-field calculations
for the Jaynes-Cummings model
\cite{Jaynes1963,Mavrogordatos2016,Shore1993,Carmichael2015}.  We also
study a highly correlated model describing Rydberg polaritons in a
cavity field. There, we introduce additional variational parameters to
also implement correlation between different modes. Finally, we give
an outlook how to extend our approach to incorporate other
nonclassical states.
        
\section{Variational Method}

	Our method is based on the idea to use the P representation of the density matrix operator to express an variational state for a bosonic field for a minimazation process. First, we want to give a brief introduction to both concepts and then we show how to combine them.

        In the context of open quantum systems, states are commonly described in terms of their density operator $\rho$. The Markovian dynamics of such open quantum systems can then be described by the Lindblad master equation $\frac{d}{dt}\rho=\mathcal{L}\rho$, with the Liouvillian $\mathcal{L}$ being the generator of the dynamics for the density matrix $\rho$ \cite{Lindblad1976} ,i.e.
	\begin{equation}
	\frac{d}{dt}\rho(t)=\mathcal{L}(\rho)=-i [H(t),\rho(t)]+\sum_i \left(c_i\rho c_i^\dagger-\frac{1}{2}\{c_i^\dagger c_i,\rho\}\right).
	\label{eq:master}
	\end{equation}
        At the same time, the equation of motions can also be expressed in the Heisenberg picture, i.e., for any given obserable $A$ as 
        \begin{equation}
	\frac{d}{dt}\hat{A}(t)=i [H,\hat{A}]+\sum_i \left(c_i \hat{A} c_i^\dagger-\frac{1}{2}{c_i^\dagger c_i,\hat{A}}\}\right).
	\label{eq:master2}
	\end{equation}
	In both equations, the jump operators $c_i$ correspond to incoherent processes, e.g. dephasing or dissipation, between the system and the bath.
        
	Solving Eq.~\eqref{eq:master} is often very challenging \cite{Weimer2020}, i.e., typically some approximations have to be made.
        Within the variational principle for open quantum systems \cite{Weimer2015}, the density matrix is approximated by the usage of a variational ansatz. In case we want to solve for the steady state of the system ($\dot{\rho}=0$) we need to minimize
	\begin{equation}
	||\dot{\rho}_\text{var}||=||\mathcal{L}(\rho_\text{var})||\rightarrow \text{min}
	\label{eq:var_master}
	\end{equation}
	with $||\dot{\rho}||=\tr[]{|\dot{\rho}|}$ being the trace norm. Here, we want to define a similar approach in the Heisenberg picture, where the steady state is defined as $\frac{d}{dt}\braket{\hat{A}}(t)=0$ for all operators $\hat{A}$. Therefore, we determine the steady state by minimizing a suitable norm for a small subset of operators given by
	\begin{equation}
	\sum\limits_n|\frac{d}{dt}\braket{\hat{A_n}}^\text{var}(t)|=\sum\limits_n|\mathcal{L}(\braket{\hat{A_n}}^\text{var})|\rightarrow \text{min}.
	\label{eq:Eom_var}
	\end{equation}
	with $\braket{\hat{A_n}}^\text{var}$ being the $n$th variational expectation value $\tr[]{\rho_\text{var} \hat{A_n}}=\braket{\hat{A_n}}^\text{var}$. 
        In the next paragraph, we will discuss how the P representation can be used to construct $\rho_\text{var}$ for bosonic fields.
	
	The most straight forward implementation of a variational density matrix is to use each entry of the matrix as a variational parameter. This is, however, not feasible in most cases. In the case of an infinite Hilbert space the number of variational paramters also goes to infinity. A solution for this problem can be found in the P representation of the density matrix \cite{Glauber1963}
	\begin{equation}
	\rho=\int d^2 \alpha\ P(\alpha) \ket{\alpha}\bra{\alpha}.
	\label{eq:coherent_rho}
	\end{equation}
	The non-orthogonality of coherent states form an overcomplete basis set of states which we can use to represent the density matrix if we combine them with an appropriate choice of quasiprobability distribution $P(\alpha)$. Such a distribution can theoretically be found for any kind of density matrix \cite{Carmichael2002} if we allow the class of generalized function in our distribution. This excludes any interpretation as an analogue to classical distribution functions because $P(\alpha)$ can become negative or more singular than a Dirac delta function $\delta(x)$.
        
	A useful property of this specific representation is the way how expectation values of annihaltion (creation) operators $a$ ($a^\dagger$) are calculated through $c$-number integrals 
	\begin{equation}
	\braket{:a^{\dagger p}a^q:}=\tr[]{\rho a^{\dagger p}a^q}=\int d^2 \alpha P(\alpha) \alpha^{*p}\alpha^q,
	\label{eq:EV_P}
	\end{equation}
	where $\braket{:a^{\dagger p}a^q:}$ indicates normal ordering of the operators. We will drop the indicator for normal ordering and always assume that the expectation values are in that order for the rest of the paper.
        
	We also want to consider the combination of different quantum states to expand the variational manifold. The construction of the corresponding P distribution is done by a convolution
	\begin{equation}
	P(\alpha)=(P_i*P_j)(\alpha)=\int d\alpha d\alpha'  P_i(\alpha') P_j(\alpha-\alpha')
	\label{eq:convolution}
	\end{equation}
	of the original distributions $P_i$ and $P_j$ \cite{Bonifacio1966}. If we insert \eqref{eq:convolution} into \eqref{eq:EV_P} we obtain
	\begin{equation}
	\braket{a^{\dagger p}a^q}=\sum_n^p \sum_m^q \xi_{p,q} \braket{(a^{\dagger})^{ n} a^{m}}_{P_i}\braket{(a^{\dagger})^{ p-n} a^{q-m}}_{P_j} 
	\label{eq:ev_formula}
	\end{equation}
	to compute expectation values of a convoluted P distribution with $\xi_{n,m}$ as the number of possible combinations of the given expectation values from $\braket{a^{\dagger n}a^m}$. For example, if we assume that one of the distribution are for the thermal state we regain the same result as in \cite{Marian1996}. We see that the calculation depends on all expectation values up to the orders $p,q$ of $a$,$a^\dagger$ of the orignial expectation value but are calculated for the single P distributions $P_i$ and $P_j$. This process can then be repeated multiple times to combine multiple distributions. Table~\ref{tab:conv1} in App.~A shows examples of the convolution of two different states.

        These ingredients are all that is required to formulate the variational principle in terms of the P distribution. The equations of motion created by
	Eq. \eqref{eq:master2} for the expectation value $\hat{A}$ depend only on expectation values like 
	\begin{equation}
	\frac{d}{dt}\braket{\hat{A}}(t)=F(\{\braket{a^{\dagger p}a^q}\}_{p,q}).
	\label{eq:eom}
	\end{equation}
	This means that we can write \eref{eq:Eom_var} as
	\begin{equation}
	D=\sum_i F_i(\{\braket{a^{\dagger p}a^q}_{p,q}\}) \rightarrow \text{min}
	\label{eq:min1}
	\end{equation}
	with $F_i$ describing the right hand side of \eref{eq:master2}, which depends on the set of expectation values $\{\braket{a^{\dagger p}a^q}\}_{p,q}$.
        To see how the P representation can be used to describe these expectation values in terms of variational paramaters, it is instructive  to have a look at some well known cases for $P(\alpha)$.
	First, we consider two classical states, a coherent and a thermal state, represented by
	\begin{align}
	&P_\text{coherent}(\alpha)=\delta(\alpha-\alpha_0)\\
	&P_\text{thermal}(\alpha)=\frac{1}{\pi n_0} \exp{\left(-\frac{|\alpha|^2}{n_0}\right)}.
	\end{align}
	
	In addition, we can find a expression for a highly non-classical state in form of the Fock states that look like
	\begin{equation}
	P_\text{fock}=\frac{1}{l!}e^{|\alpha|^2} \frac{\partial^{2l}}{\partial \alpha^l \partial \alpha^{*l}}\delta^{(2)}(\alpha)
	\end{equation}
	The distribution includes derivatives of the delta distribution which are defined as 
	$\int dx \delta^{(n)}(x) \psi(x)=(-1)^n \psi^{(n)}(0)$. 
	We can immediately see that each distribution has one defining parameter, i.e., $\alpha_0\in \mathbb{C}$, $n_0\in \mathbb{R}$, or $l \in \mathbb{N}$. The convolution of two P distributions results in a new P-distribution that depends on the set $\{\beta\}=\alpha_0,n_0,l$.
	This allows us to formulate Eq. \eqref{eq:min1} as 
	\begin{equation}
	D=\sum_i F_i(\{\beta\}) \rightarrow \text{min}
	\label{eq:min2}
	\end{equation}
	Upon inspection of Eq. \eqref{eq:ev_formula} we can also see that we do not need to know the complete form of the P distribution that corresponds to a specific state. Instead, it is enough to know how all expectation values depend on the variational parameters $\beta$. This is for example useful if it is difficult to find a complete expression of the P distribution like in the case of the squeezed coherent state. The state can be obtained by convolution of the coherent state and the squeezed vacuum state where the distribution is known \cite{Kiesel2009} or by directly evaluating the expectation values for this particular state. In this case we know that the squeezing operator $S^{(\dagger)}(r,\Phi)$ with squeezing parameter $r$ and angle $\Phi$ changes the annihaltion operator $\hat{a}$ like 
	\begin{equation}
	S^{\dagger} \hat{a} S=
	\hat{a} \cosh(r)-e^{i\Phi} \hat{a}^\dagger \sinh(r)
	\end{equation} 
	which allows us to directly calculate how the expectation values in \eref{eq:min1} depend on the parameters $r$ and $\Phi$.
	
	\section{Jaynes-Cummings Model}

        In order to benchmark our variational approach, let us turn to a driven-dissipative variant of the Jaynes-Cummings model, where we compare the variational method to WFMC simulations using the QuTiP package  \cite{Johansson2012,Johansson2013}. The Jaynes-Cummings model describes a atom interacting with a light field that is trapped inside a cavity. The Hamiltonian is of the form
	\begin{equation}
	H=\Delta_c a^\dagger a +\Delta_a\sigma^+\sigma^-+g(a\sigma^+ +a^\dagger \sigma^-)+p(a^\dagger+a).
	\end{equation} 
	The first two terms describe the detunings $\Delta_c$,  $\Delta_a$ for the cavity respectivly the atoms from the driving frequency. The atom and the cavity are coupled with a strength $g$ and we pump the cavity with an driving amplitude $p$. Additionally, we include cavity losses and spontaneous emission of the atoms into other modes than the cavity via the jump operators $c_c=\sqrt{\gamma} a$ and $c_a=\sqrt{\kappa} \sigma^-$ with decay rate $\gamma$ for the cavity mode and $\kappa$ for the atom, respectively.

        We use a product ansatz for the atom and the cavity
	\begin{equation}
	\rho=\rho_\text{cavity}\otimes \rho_\text{atom}
	\label{eq:product_states}
	\end{equation}
	in the variational approach and use the variational parameter $\alpha_i$ in $\sum_{i=0,x,y,z} \alpha_i \sigma_i$ to describe the atomic part, while we use the P representation to account for the cavity mode. As our variational parameter set we use a convolution of coherent, thermal, fock and squeezed states.

\begin{figure}
		\includegraphics[width=8.6cm]{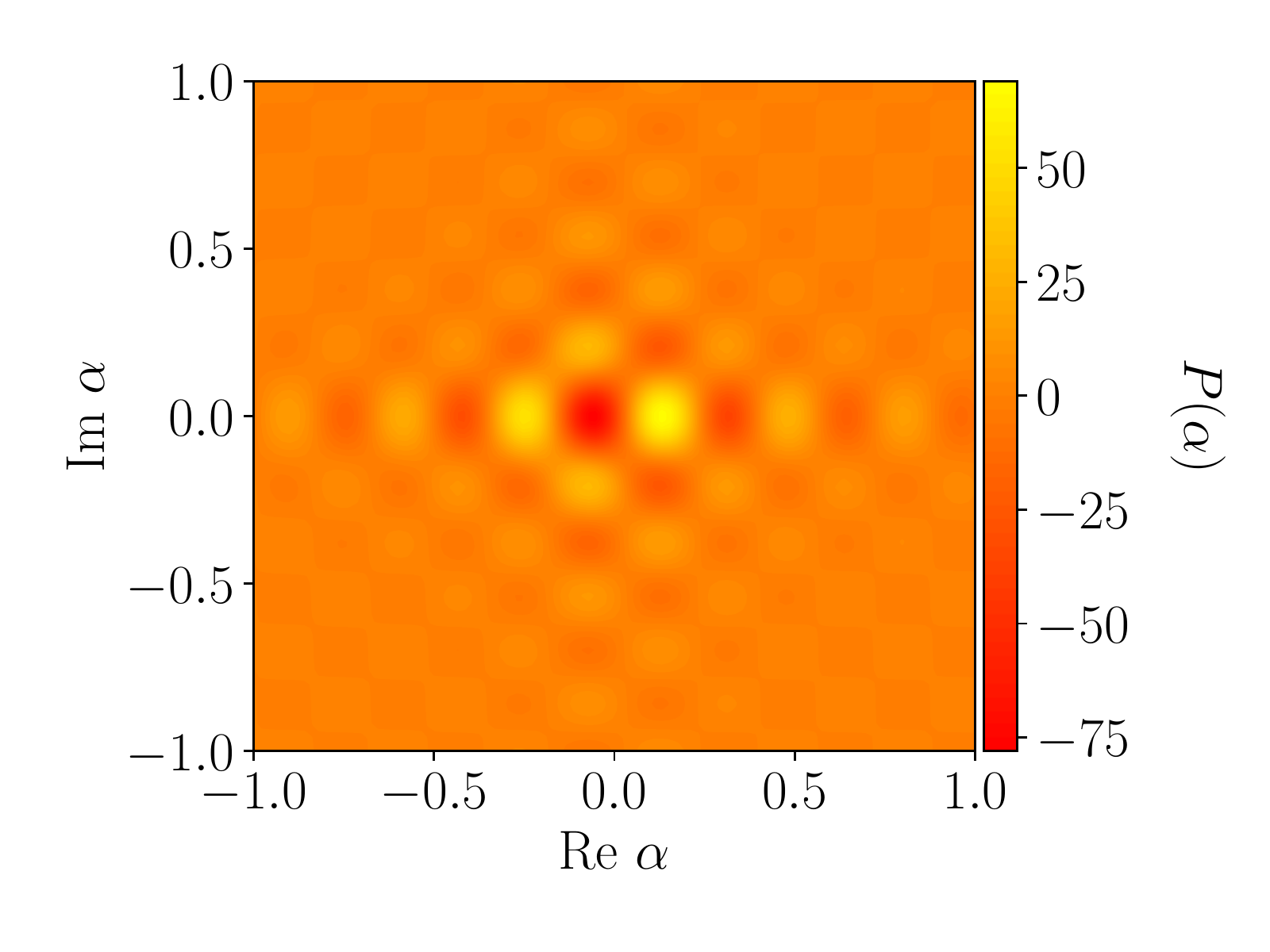}
		\caption{P-distribution of the Jaynes-Cummings model for $g/\gamma=3347$, $\kappa/\gamma=6$,$\Delta_c/\gamma=340$,$\Delta_a/\gamma=23.5\times 10^3$ and $p/\gamma=50$. The distribution shows regions of strong negativity which indicates a non-classical behavior of the model for the given parameters. The distribution is obtained via the variational principle.}
		\label{fig:P_dis}
	\end{figure}

        To show an immediate advantage of the variational approach, we also want to analyse the Maxwell-Bloch equations of the Jaynes-Cummings model \cite{Carmichael2015,Mavrogordatos2016}. This set of equation describe the time evolution of the lowest order of expectation values. The atom and cavity decouples in a similar fashion like in \eref{eq:product_states}, but it also decouples the equation from higher order terms of the cavity field through the neglection of any correlation term of the second or higher order \cite{Castro2015}. The Maxwell-Bloch equations for the Jaynes-Cummings model are given by
	\begin{align}
	\frac{d}{dt}\braket{a}&=-(\kappa+i\delta_c)\braket{a}-ig\braket{\sigma^-}-ip\\
	\frac{d}{dt}\braket{\sigma^-}&=-(\frac{\gamma}{2}-i\Delta_a)\braket{\sigma^-}+ig\braket{a}\braket{\sigma^z}\\
	\frac{d}{dt}\braket{\sigma^z}&=-\gamma(\braket{\sigma^z}+1)+2ig(\braket{a^\dagger}\braket{\sigma^-}-\braket{a}\braket{\sigma^+}).
	\end{align}
	\begin{figure}
	\includegraphics[width=8.6cm]{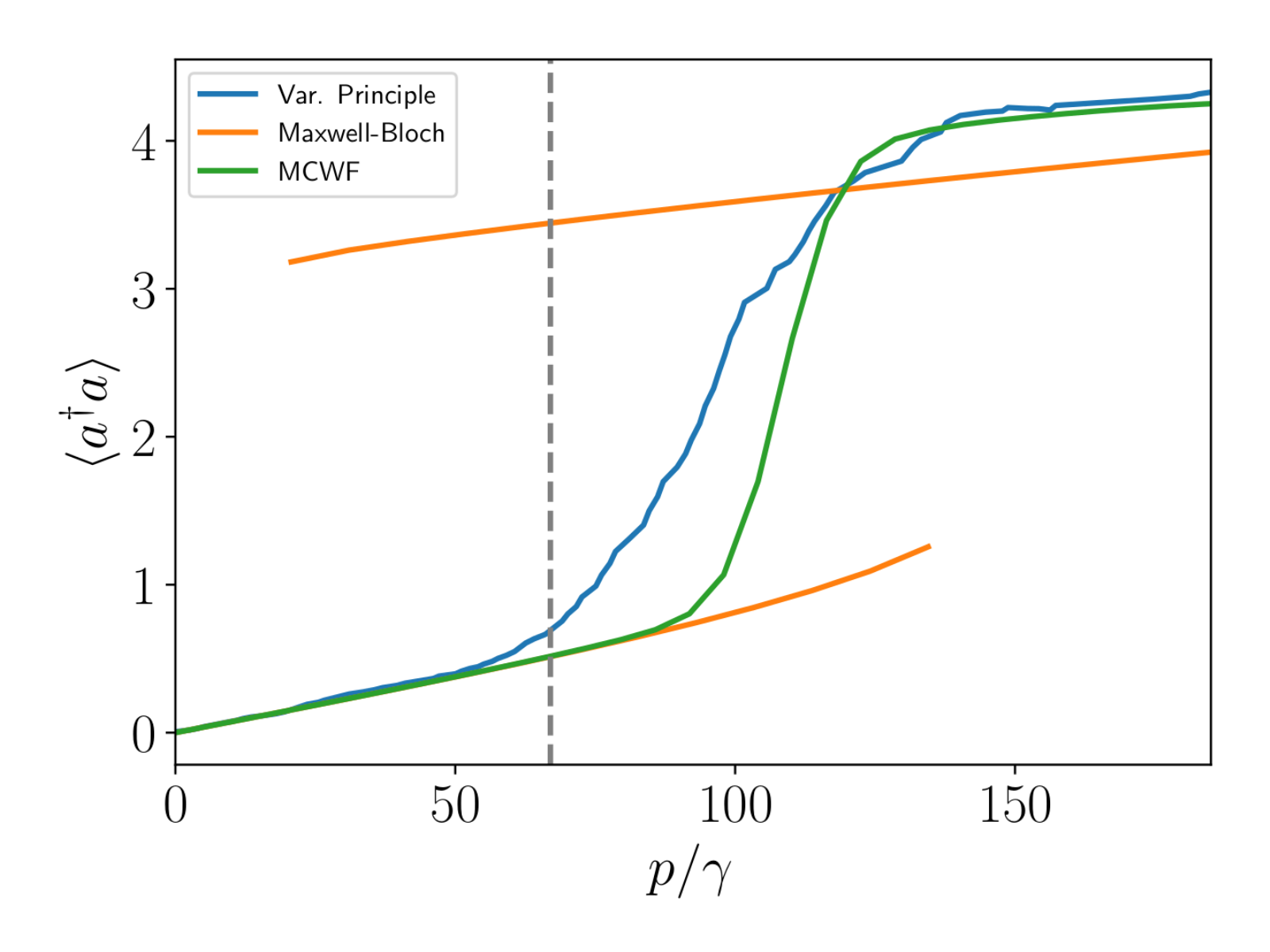}
	\caption{Results of  the Maxwell-Bloch equations, the variational approach and the Monte-Carlo wavefunction method (MCWF) for the Jaynes-Cummings model with  $g/\gamma=3347$, $\kappa/\gamma=6$,$\Delta_c/\gamma=340$,$\Delta_a/\gamma=23.5\times 10^3$. The Maxwell-Bloch equations (orange) show a region of bistability which can be solved by a variational treatment of the equations which yields a prediction for the transition between the two solution at the grey line. For higher orders of the variational approach (blue) shifts the transition towards the MCWF solution.}
	\label{fig:jc_int}
\end{figure}
	Fig. \ref{fig:jc_int} shows a comparison between the solution
        of the Maxwell-Bloch equations, the Monte-Carlo wave function
        solution and the variational expectation value approach for
        the cavity field $\braket{a^\dagger a}$. The mean-field
        solution (orange) shows a large area of bistability between
        two solutions. A comparison of the norms of the two solutions inserted in a set of first-order equations of motion resolves the bistability and indicates a jump between the solutions at the grey line. The third line (blue) shows the solution of expanding the equations of motion up to second order in the variational approach, resulting in a clear improvement. Fig. \ref{fig:P_dis} shows a reconstructed P distribution from the variational expectation values through the usage of the characteristic function  
	\begin{equation}
	\chi(z)=\sum\limits^\infty_{k,l=0}\frac{z^k}{k!}\frac{(-z^*)^l}{l!}\braket{a^{\dagger k}a^l}
	\end{equation}
	and 
	\begin{equation}
	P(\alpha,\alpha^*)=\frac{1}{\pi^2} \int\limits_{-\infty}^{\infty} d^2 z \chi(z) e^{-iz^*a^\dagger}e^{-iza}.
	\end{equation}
        The non-classicality of the steady state is clearly shown by
        the negative values of $P(\alpha)$. The remaining difference
        with the WFMC simulations can be attributed to the neglection
        of correlations between the atom and the cavity mode due to
        our product ansatz in Eq.~(\ref{eq:product_states}).
        
        

	%
	%

	\section{Rydberg cavity polaritons}

        Let us now turn to a model where correlations beyond a single
        mode are particularly important. For this, we investigate an
        effective three-boson model to describe strongly interacting
        Rydberg atoms inside a cavity \cite{Grankin2014,Grankin2015},
        which describes nonlinear effects that arises from the
        interaction of the Rydberg atoms. Before turning to the
        variational analysis, we briefly want to recapture the key
        pieces of the model.

        Consider a cavity filled with $N$ three-level atoms with energy level $g,e,r$ as the ground state $\ket{g}$, an intermediate level $\ket{e}$ and an highly excited state which we denote as the Rydberg level $\ket{r}$. The key idea is to restrict the dynamics to three bosonic modes that describe the cavity mode and the symmetric subspaces of the atomic excitations. This restriction of the atoms to their symmetric subspace is valid as long as the total number of atomic excitations is small compared to $N$ \cite{Grankin2015}.
     
	\begin{figure}
	\centering
	\includegraphics[width=1.\linewidth]{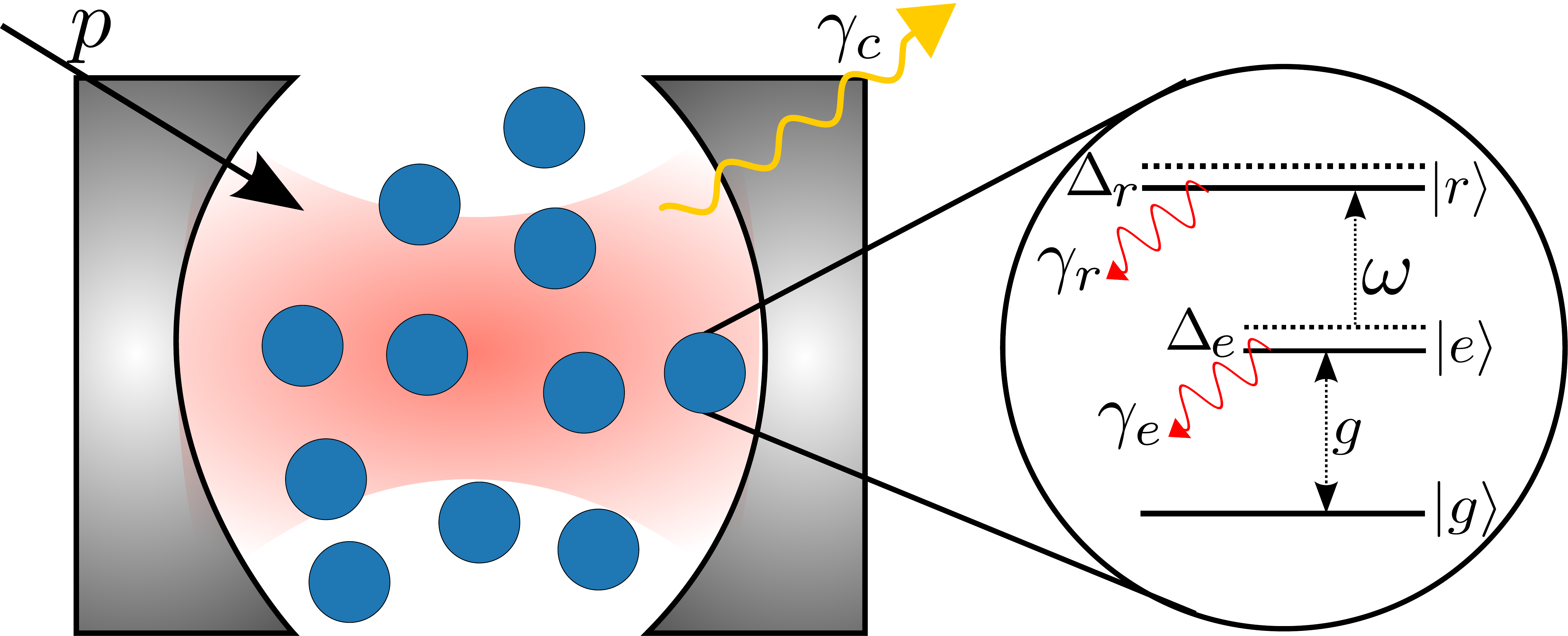}
	\caption{Scheme of multiple Rydberg atoms trapped inside a cavity. The atoms are described by a three-level ladder scheme
	with $g$ as the coupling constant between the cavity light field and the transition between the
	ground state $\ket{g}$ and the intermediate state $\ket{e}$. A control laser then
	couples the intermediate to the Rydberg state $\ket{r}$ with a strength
	of $\omega$. The one- and two-photon detunings of the atoms are given by $\Delta_e$ and $\Delta_r$. The rates $\gamma_c$, $\gamma_e$ and $\gamma_r$ describe decay processes of the cavity, the intermediate atomic state and the Rydberg state.}
	\label{fig:three_mode_model_2}
	\end{figure}

	We then can describe the system in terms of collective operators describing the symmetric subspace with $a$ being the annihilation operator for the cavity mode and  $b=\frac{1}{\sqrt{N} }\sum\limits_{n=1}^N \sigma_{ge}$ and $c=\frac{1}{\sqrt{N} }\sum\limits_{n=1}^N \sigma_{gr}$ as the collective operators for the atomic transition modes $\sigma_{ge}$ and $ \sigma_{gr}$.
	With that the Hamiltonian reads as
	\begin{align}
	H&=-\Delta_c a^\dagger a+ p(a+a^\dagger)-\Delta_e b^\dagger b - \Delta_r c^\dagger c \nonumber\\
	& +g\sqrt{N} (ab^\dagger+a^\dagger b)+\frac{\omega}{2}(bc^\dagger+b^\dagger c)+\frac{\kappa_r}{2} c^\dagger c^\dagger cc
	\label{eq:h_eff}
	\end{align} 
	and the the jump operators are given by $c_e=\sqrt{\gamma_e} b$, $c_r=\sqrt{\gamma_r} c$ for the intermediate and Rydberg state and also for the cavity. The nonlinear terms in the Hamiltonian arise from the van-der-Waals-force between atoms in the Rydberg state. The interaction also couples the symmetric subspace to the antisymmetric subspace which leads to an additional nonlinear dissipation term  $c_{nl}=\sqrt{\kappa_i} cc$.
        
	We now study the model by working in the eigenbasis of the non-interacting Hamiltonian at $\kappa_r=0$. The diagonalisation of Eq.\eqref{eq:h_eff} results in $H=\sum_{q\in +,0,-} c_q \Psi_q^\dagger \Psi_q$. The new states $\Psi_q$ form polariton states. There are defined as a quasi particle consisting of both light and matter. For a three level atomic system we get two different types of polaritons. The ones with $q \in \pm$ are bright state polaritons, while $q=0$ is the dark state polariton. The dark state polariton shows very different behavior as it is decoupled from the intermediate atomic level, which leads to long lifetimes in the cavity.

        The previously neglected interaction between the polariton leads to a strongly correlated many-body system which provides a difficult task for numerical calculation especially for large atom numbers \cite{Pistorius2020}.
        
	To also be able to capture correlations between the modes we need additional variational parameters. If we look at the lowest order expectation values between different modes we get
	\begin{equation}
	\braket{ab}=\braket{a}\braket{b}+\delta(ab)
	\label{eq:cf}
	\end{equation}
	with $\delta(ab)$ as the correlation function between mode $a$ and $b$.  These kind of factorizations for expectation values can be done for all orders and provides us with the needed variational parameter in form of the correlation functions $\delta(a^n b^m)$ \cite{Leymann2013,Leymann2014,Schoeller1994}.
        
\begin{figure}
		\includegraphics[width=8.6cm]{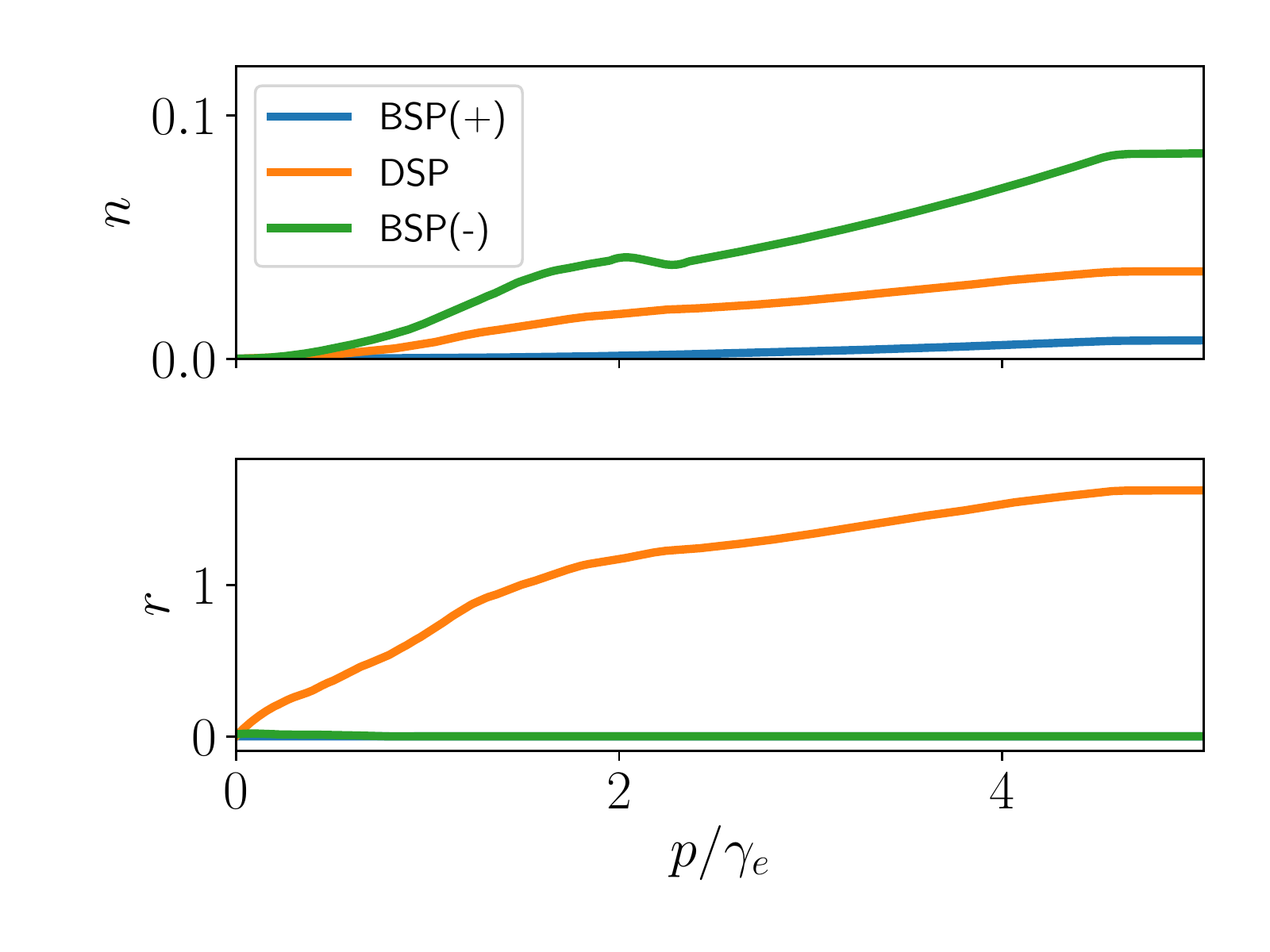}
	        \caption{Intensity(Top) and squeezing parameter $r$ (buttom) of the effective three boson model in the polariton picture for parameters: $\gamma_r=0.1\,\gamma_e$, $\gamma_c=0.3\,\gamma_e$, $\Delta_c=0$, $\Delta_e=-10\,\gamma_e$, $\Delta_r=0$, $\kappa_r=-1.2\,\gamma_e$, $\kappa_i=0.42\,\gamma_e$, $g=4.2\,\gamma_e$, $N=10^4$. The intensity of all polariton modes increase with the pumping strength and reaches a saturation at around $p\approx 5\gamma_e$. However, only the dark state polariton mode displays squeezing. For the bright state polaritons, the squeezing parameter essentially vanishes.}
	\label{fig:boson_model}
\end{figure}

	Fig. \ref{fig:boson_model} shows the occupation number of the different modes and their squeezing strength as the parameter $r$. The photons are getting absorbed by the different photon modes. The bright state polaritons show for the off-resonant parameters we have chosen here an uneven population. All modes reach a saturation around $p\approx 5\gamma_e$. If we look at squeezing parameter we can see that this is the only mode that experiences strong nonlinear effects while the squeezing is mostly suppressed for the other two.

        Our results also demonstrate the importance including
        correlations between the modes. Without them, we find that the
        occupation numbers can become very large (e.g., up to
        $n=200$), which correspond to states with very large (and hence unphysical) van der Waals interaction energies.
        
%
%
	
	\section{Possible extensions}
        
        In all of our previous calculations, we worked with only a handful of different convoluted states to successfully construct our variational manifold. However, we would like to point out that it is possible to extend our approach toeven broader classes of quantum states. As already mentioned, it is not necessary to know the full P distribution function as  it is sufficient to be able to calculate expectation values of the given state, which gives us access to a great variety of non-classical states.
        
	In the previous chapters we already discussed the coherent squeezed states as the most prominent candidate for squeezing but there are similar definitions for squeezed Fock state $\ket{l}_{sf}$ and squeezed thermal states $\ket{n}_{sth}$ \cite{Marian1991,Kim1989,Kim1989a}
	\begin{align}
	\ket{l}_{sf}&=S(r,\phi)\ket{l}\\
	\ket{n}_{sth}&=S(r,\phi)\ket{n}.
	\end{align}
	Both classes of states have already been investigated in some detail, with explicit expression for expectation values of all orders being known \cite{Marian1993,Marian1996}. Hence, these squeezed Fock states can also be readily integrated into our variational approach.

        Furthermore, it is also possible to employ highly entangled Schr\"odinger cat states given by	
	\begin{equation}
	\ket{\psi}=A(\ket{\alpha_1}+\Theta \ket{\alpha_2}),
	\end{equation}
	with $\ket{\alpha_1}$ and $\ket{\alpha_2}$ being two different coherent states. The expectation values for this state can be calculated via the explicit P distribution \cite{Brewster2018}.
        
\begin{figure}

	  \begin{tabular}{cc}
            
	    \includegraphics[width=4.2cm]{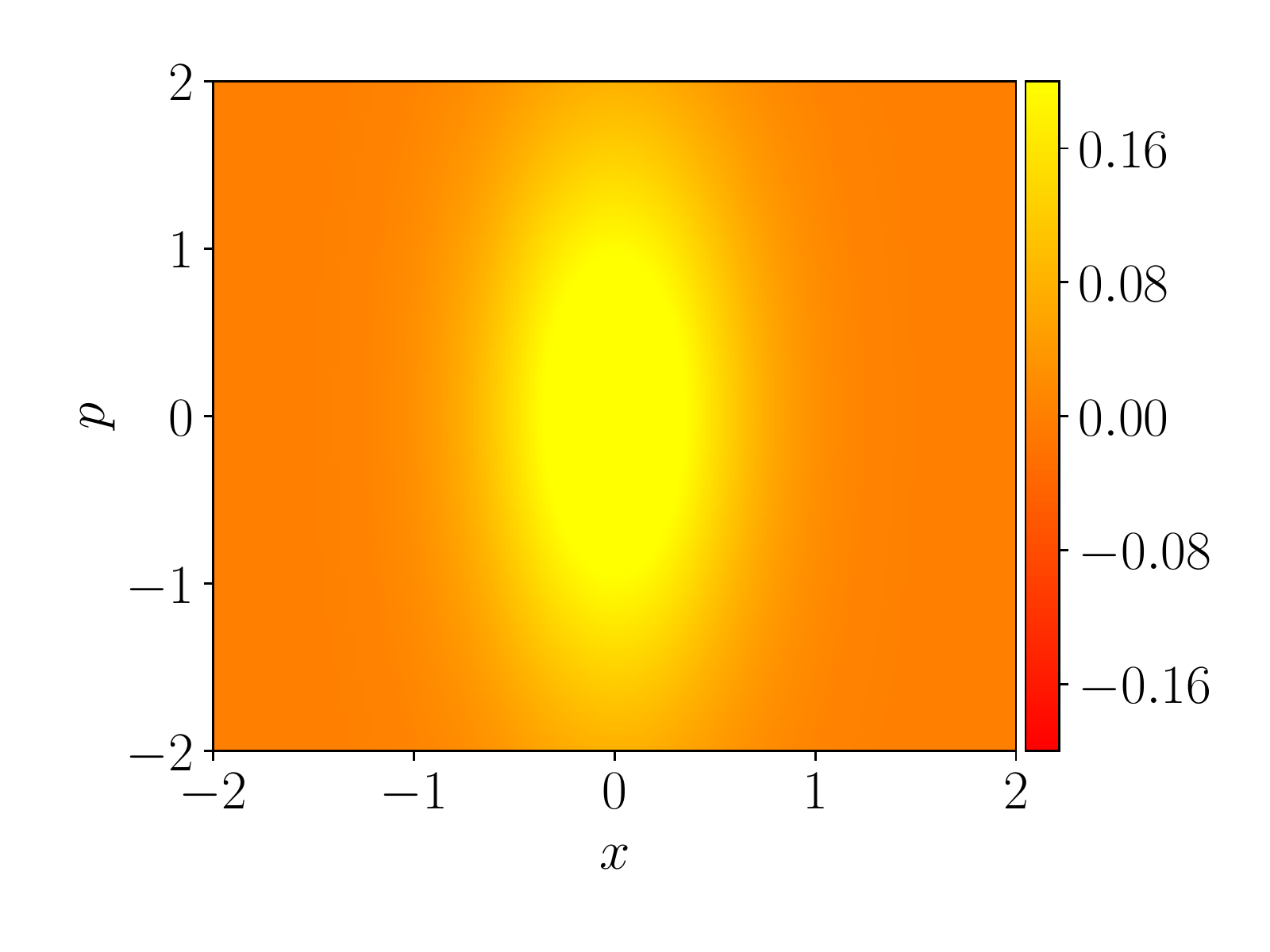} & \includegraphics[width=4.2cm]{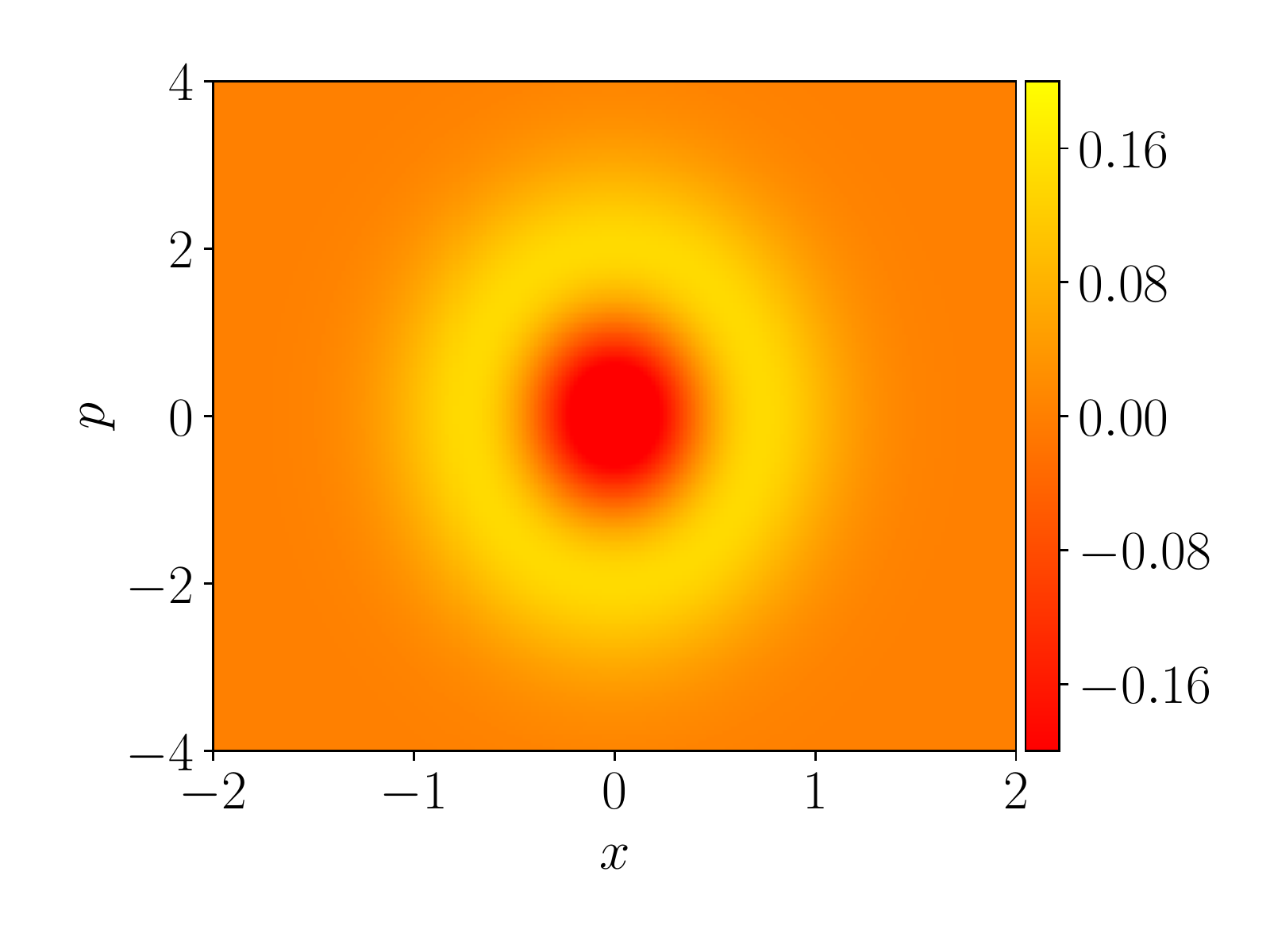}\\
            (a) & (b)
          \end{tabular}
          \begin{center}
            \includegraphics[width=4.2cm]{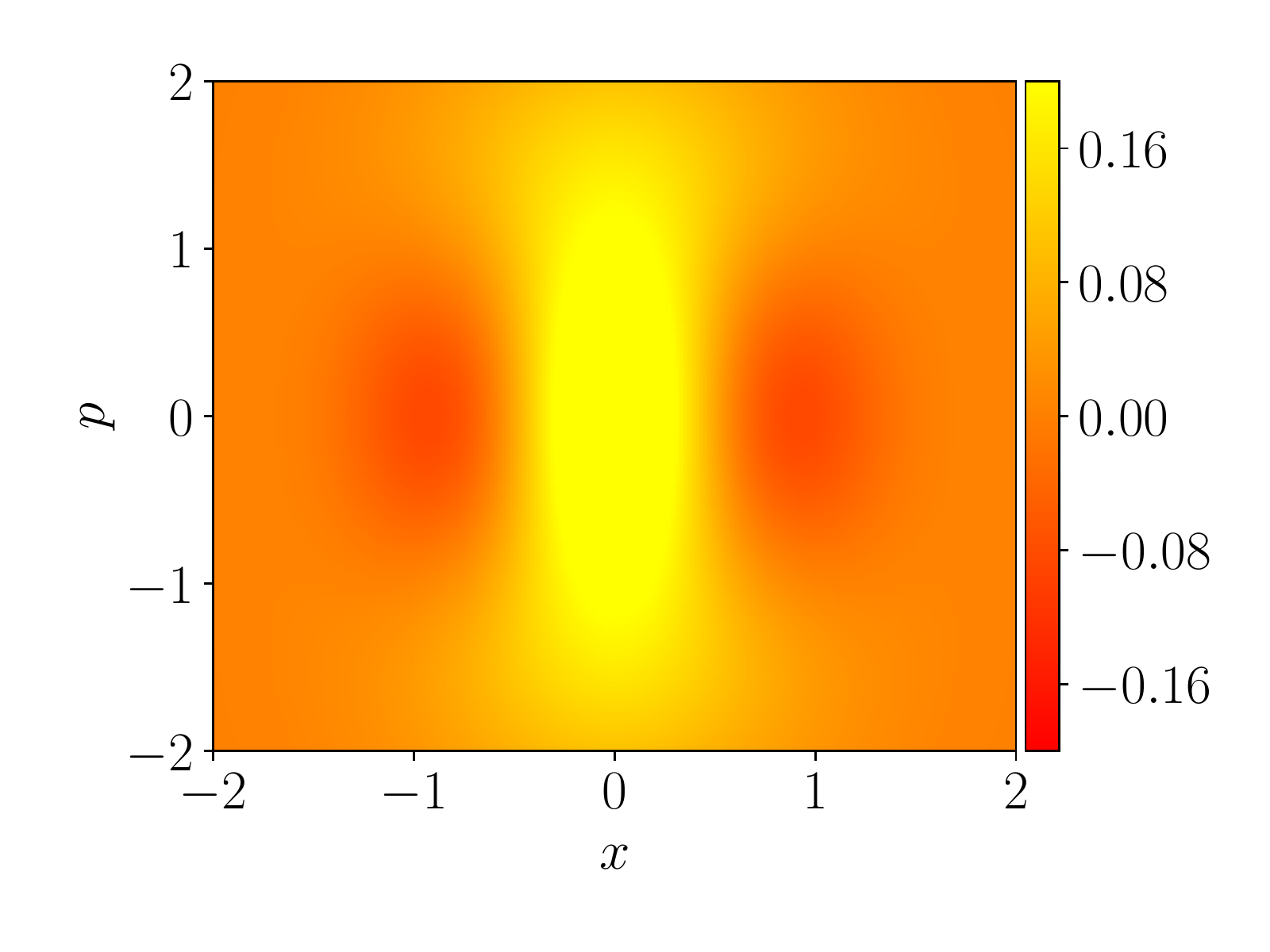}\\
            (c)
            \end{center}

	\caption{Wigner function for different nonclassical states.(a) squeezed thermal state and (b) squeezed Fock state  and (c) Schr\"odinger cat state.}	
	
	\label{fig:new_states}
\end{figure}

	Fig. \ref{fig:new_states} shows the Wigner distribution of all three states. The Wigner distribution is more suitable for a visual representation because it does not have singularities for highly non-classical states that are found in the P distribution. The transformation 
	\begin{equation}
	W(\alpha)=\frac{2}{\pi} \int d^2\alpha' e^{-2|\alpha-\alpha'|^2} P(\alpha')
	\end{equation}
	connects both distributions.

        We also want to make a clear distinction between the squeezed thermal (Fock) state and the convoluted distribution of a squeezed coherent state with a thermal (Fock) state. Especially in the case of the thermal state it is not straight forward to see from their Wigner functions that the two results are actually different. Therefore, it is instructive to look at the difference of their intensities, which is given by
        \begin{equation}
          \braket{a^\dagger a}_{sth}-\braket{a^\dagger a}_{s+th}=2n_0\sinh^2{r}.
        \end{equation}
 The difference is even enhanced for higher-order expectation values, which can significantly change the result of the minimization in Eq.~\eqref{eq:min2}.
        
 In case of the cat state the situation is reversed. Although the visual representation in \ref{fig:new_states} is cleary distinguishable from a simple coherent state, the difference enters only in higher orders, as the lowest order is given by $\braket{a}_\text{cat}=\alpha_1+\alpha_2=\tilde{\alpha}$. Only the scaling with higher order expectation values can reveal the true nature of this state and shows the importance of incorporating as many orders as possible for the equations of motion.
        
 Finally, we would like to mention two additional classes of states
 that could be included in the variational analysis. Both the single-variable Hermite polynomial states \cite{Hillery1987,Bergou1991,Tan2015,Agarwal1992} and the photon-added (substracted) coherent states \cite{Zavatta2007,Barnett2018} appear to be good candidate for a further expansion of the variational approach.

	\section{Summary}
        
	In summary, we have extended the variational principle for
        open quantum systems through the usage of the P distribution
        of the density matrix. Despite its simplicity, we find that
        our method yields even quantitatively reasonable results for
        the driven-dissipative Jaynes-Cummings model. Furthermore, we have succesfully applied our approach to an effective model describe a many-body system of Rydberg atoms in a cavity, where we can identify strong squeezing of a dark state polariton mode. Our approach could be especially fruitful for applications where strong nonclassical correlations play an important role, such as gravitational wave detection using squeezed light \cite{LIGO2011,Zhao2020,McCuller2020} or the preparation of nonclassical states of light in photonic condensates \cite{Kurtscheid2019}. Finally, we have presented several directions how the class of variational states could be extended further.

        \begin{acknowledgments}

We thank Jingtao Fan for fruitful discussions. This work was funded
by the Volkswagen Foundation, by the Deutsche Forschungsgemeinschaft
(DFG, German Research Foundation) within SFB 1227 (DQ-mat, project
A04), SPP 1929 (GiRyd), and under Germany’s Excellence Strategy --
EXC-2123 QuantumFrontiers -- 390837967.
\end{acknowledgments}

	\bibliographystyle{aip}
	\bibliography{bib_P_distribution}

	\clearpage
	\onecolumngrid
	\appendix
        \section{Convoluted P distributions}

Tab.~\ref{tab:conv1} shows examples of convolutions of two P distributions used in the main text.

	\begin{table*}[h]

	  \begin{tabular}{p{1.5cm}|p{4.cm}p{4.cm}p{4.cm}p{4.cm}}
		
		&Coherent\linebreak $\alpha_1=i$&Squeezed coherent state \vfill $r_1=0.5$,$\phi_1=0$  & Thermal state\vfill$n_1=0.1$ &Fock state\vfill$l_1=1$ \\
		\hline
		\vspace{-2.cm}Identity Matrix &\includegraphics[width=4.cm]{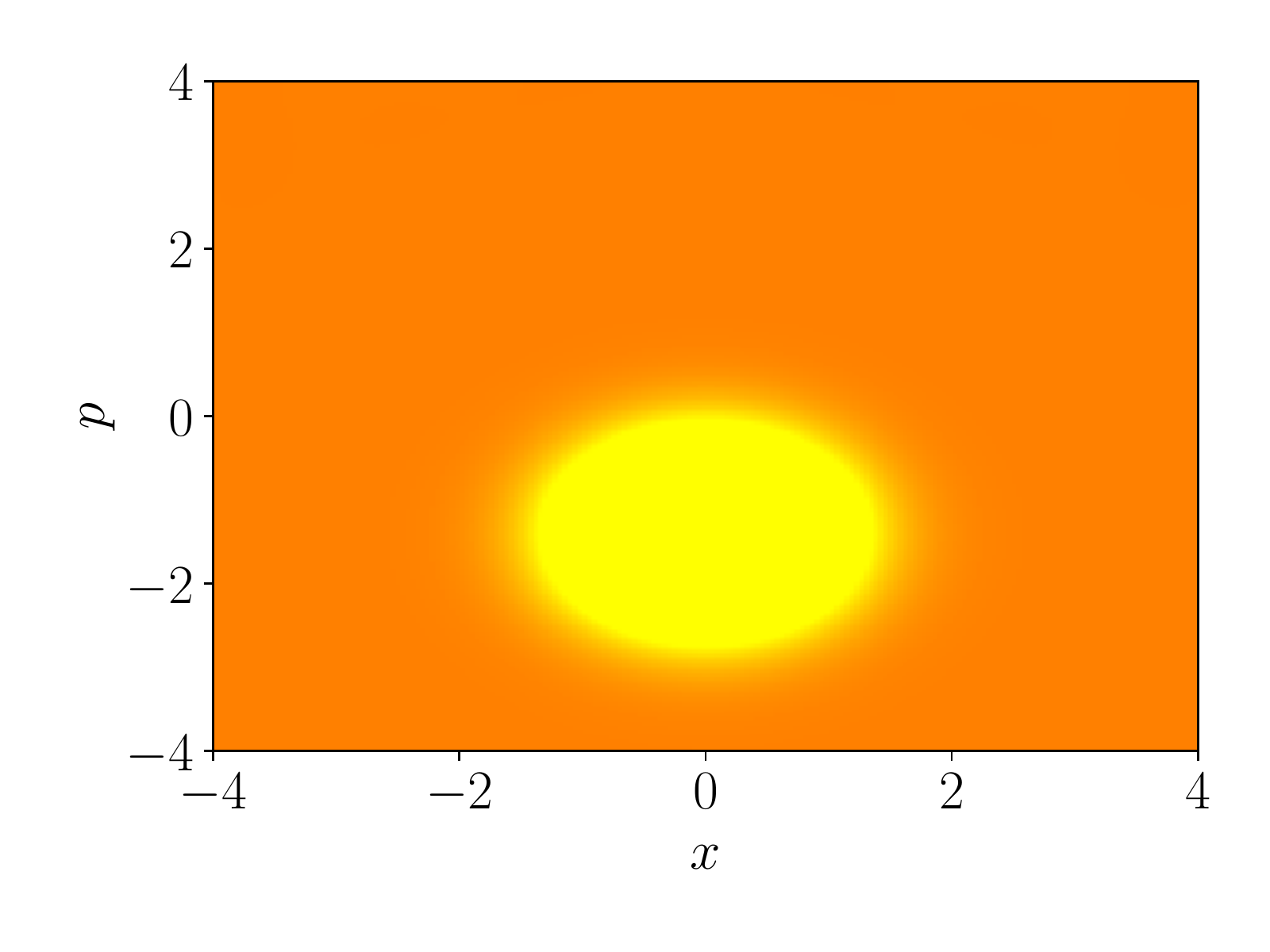} &\includegraphics[width=4.cm]{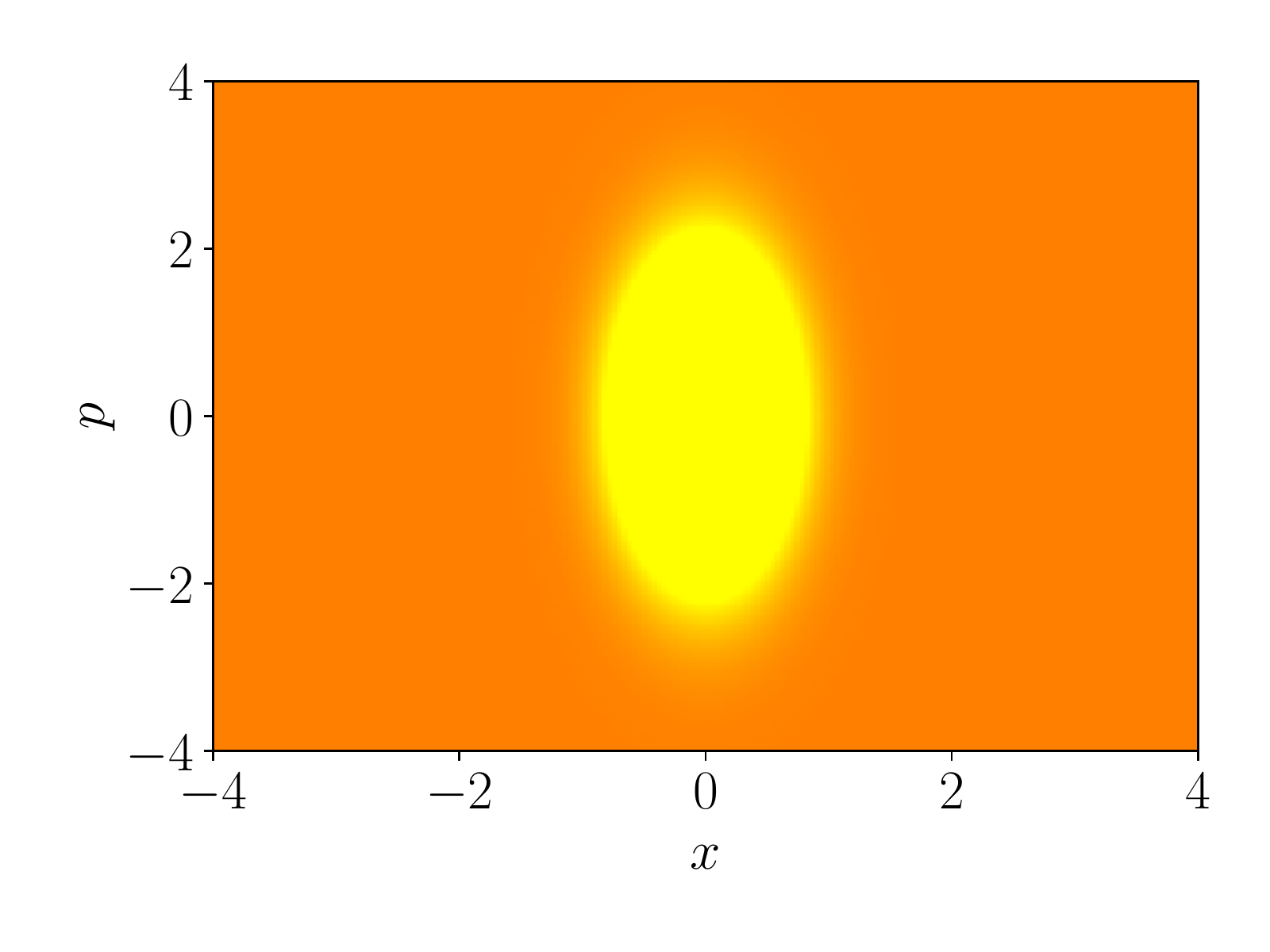} &\includegraphics[width=4.cm]{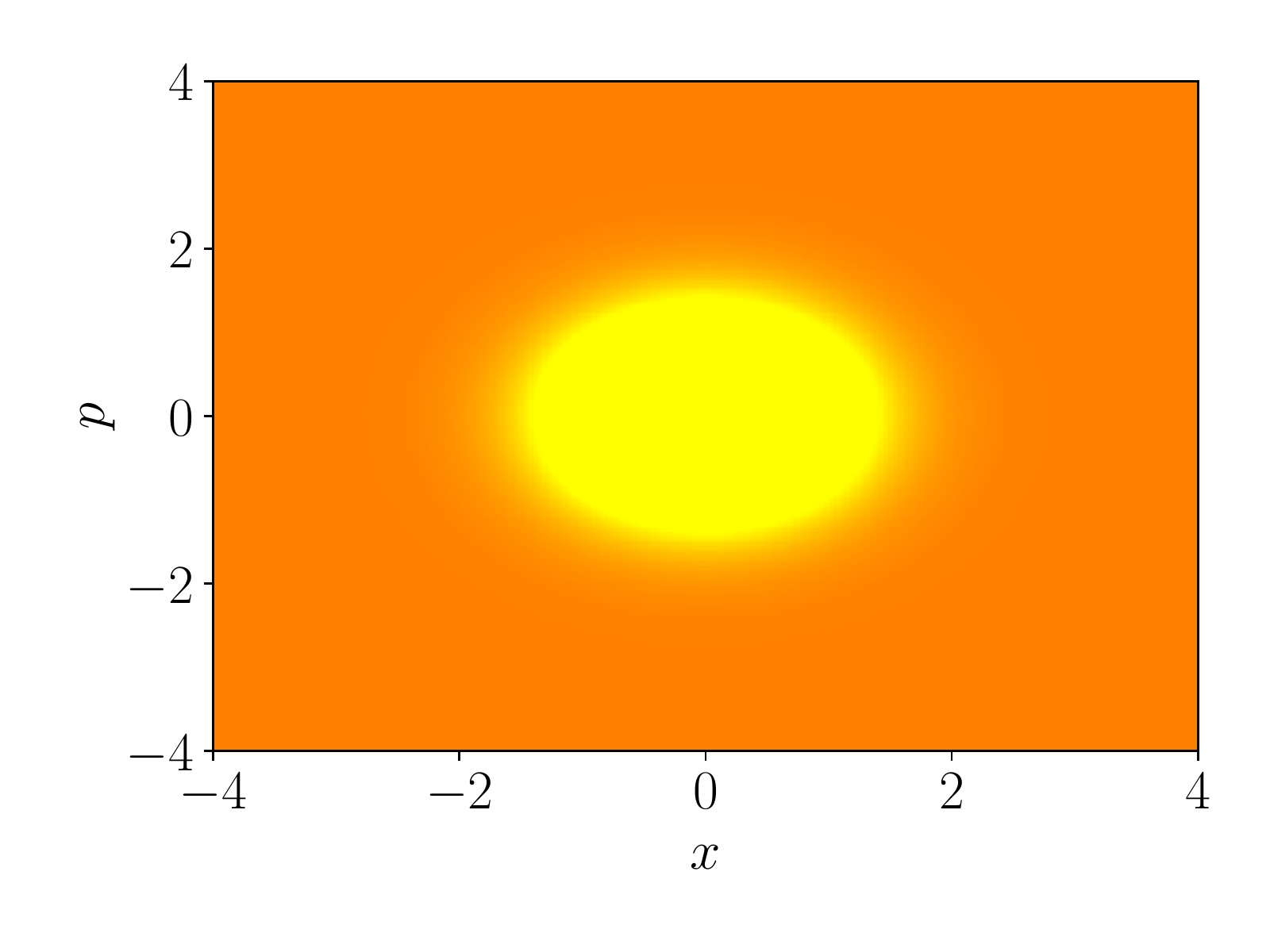} 
		&\includegraphics[width=4.cm]{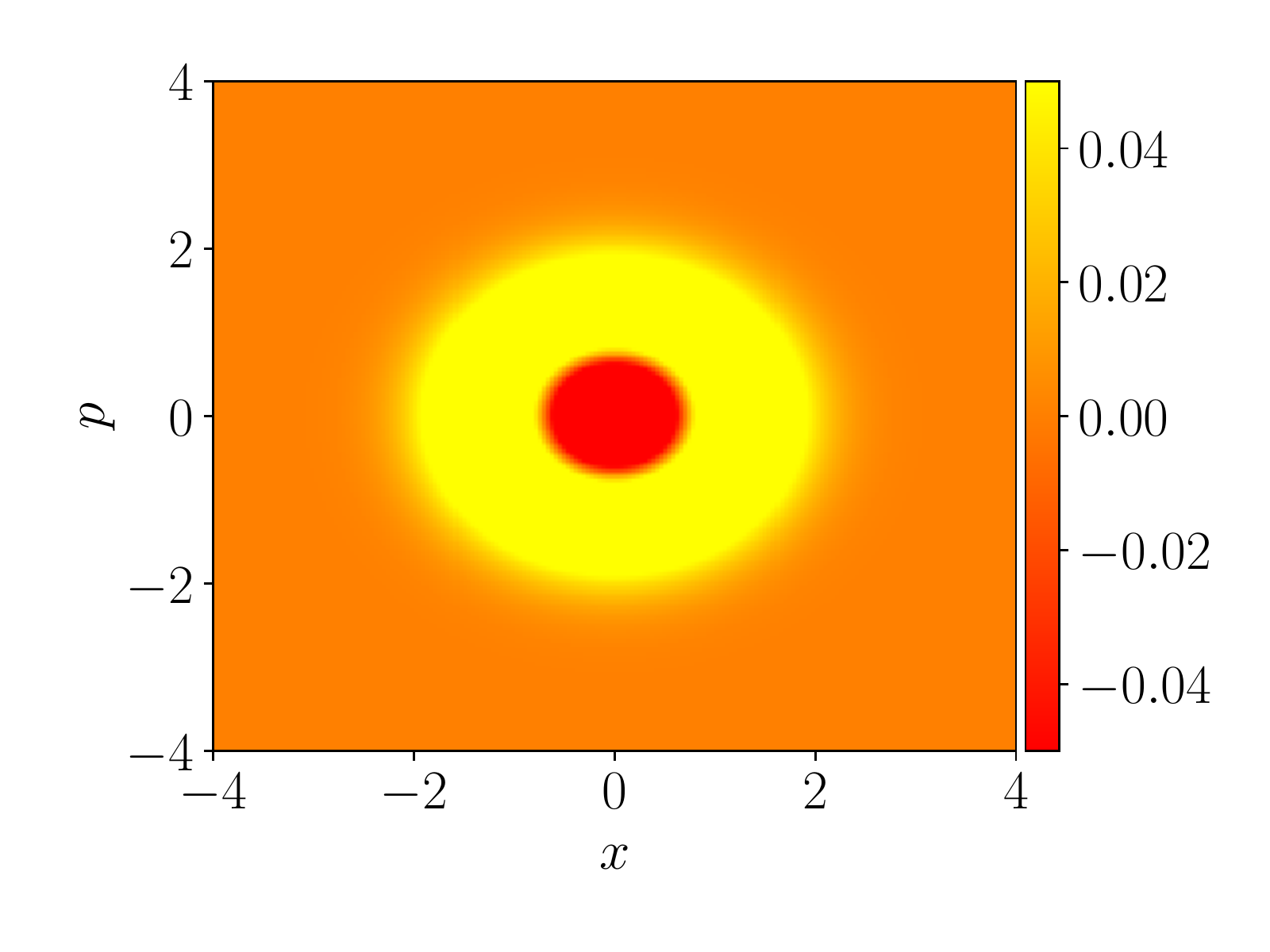} \\
		\vspace{-2.cm}Coherent state\vfill$\alpha_2=1$ &\includegraphics[width=4.cm]{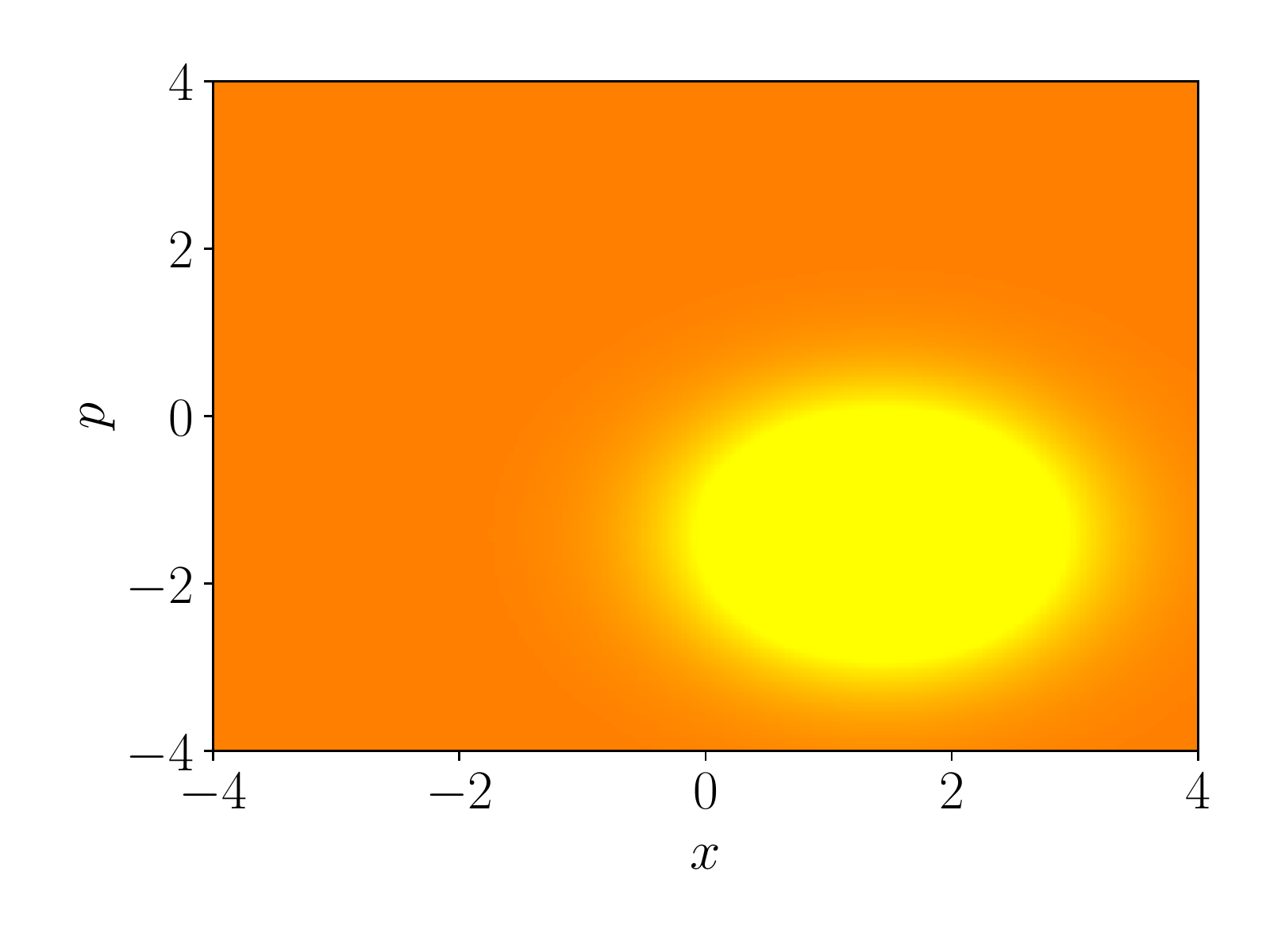} &\includegraphics[width=4.cm]{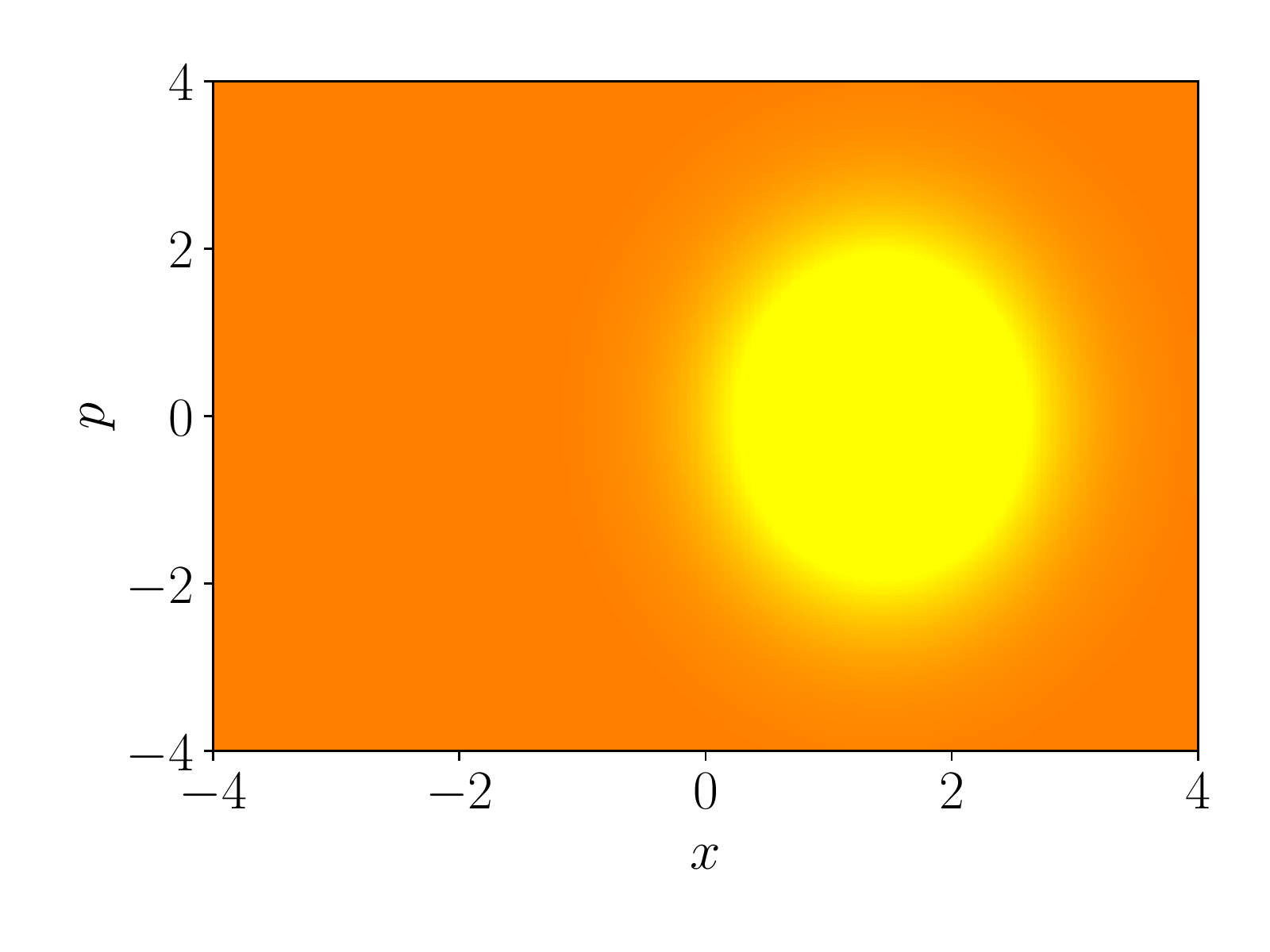} &\includegraphics[width=4.cm]{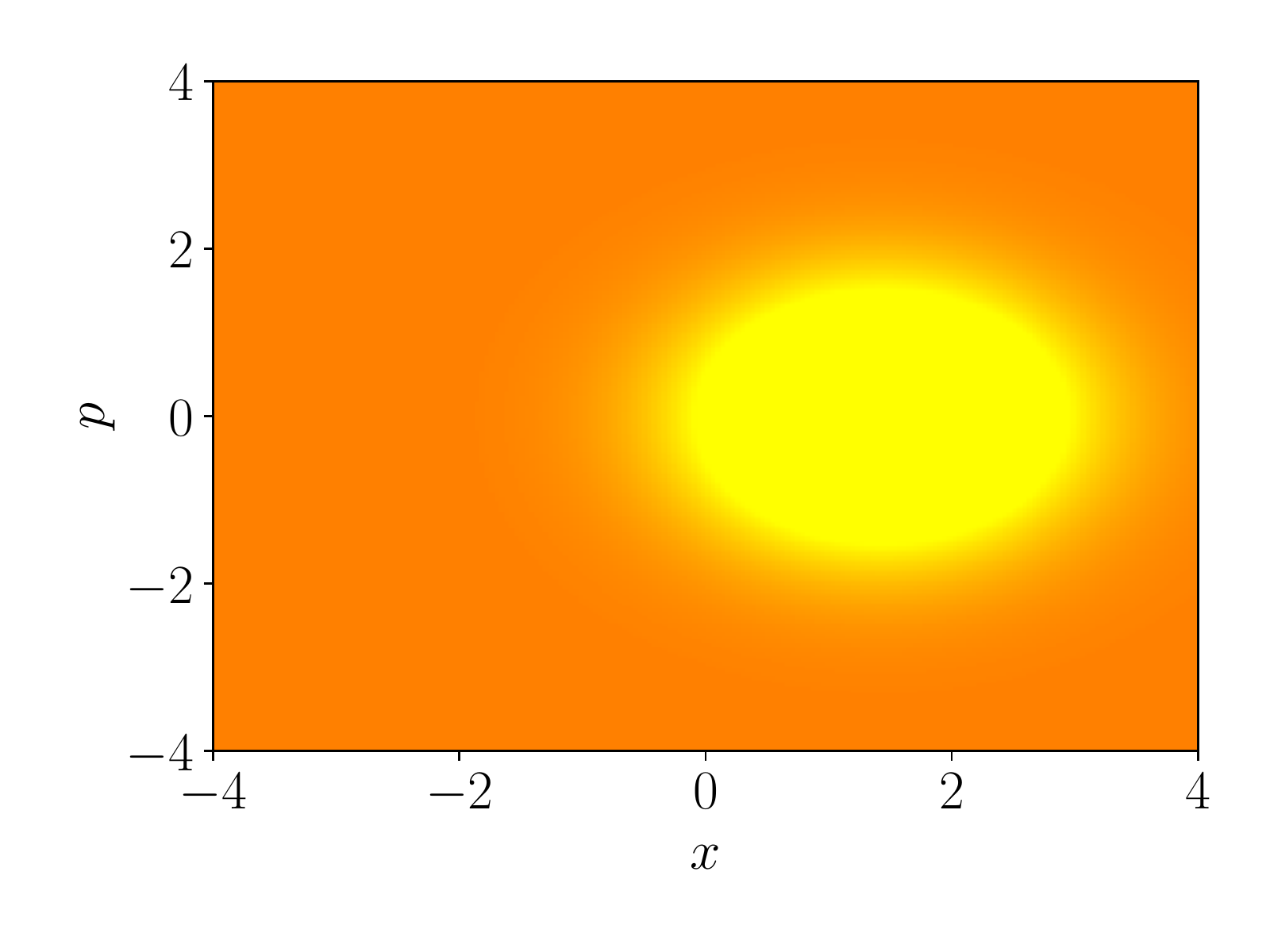} &\includegraphics[width=4.cm]{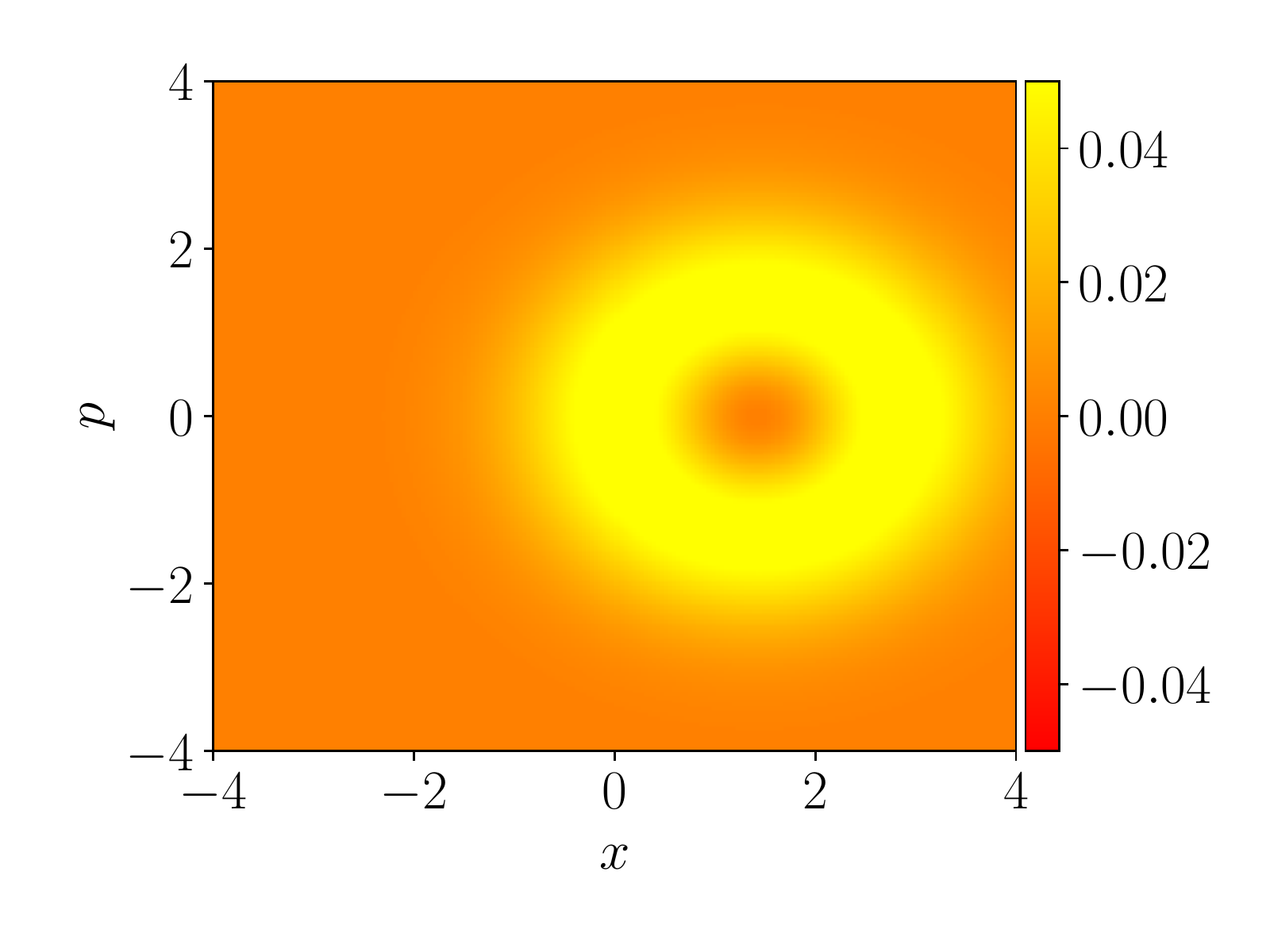}\\
		\vspace{-2.cm}Squeezed coherent state \linebreak$r_2=1,$\linebreak $\phi_2=-\pi/2$ &\includegraphics[width=4.cm]{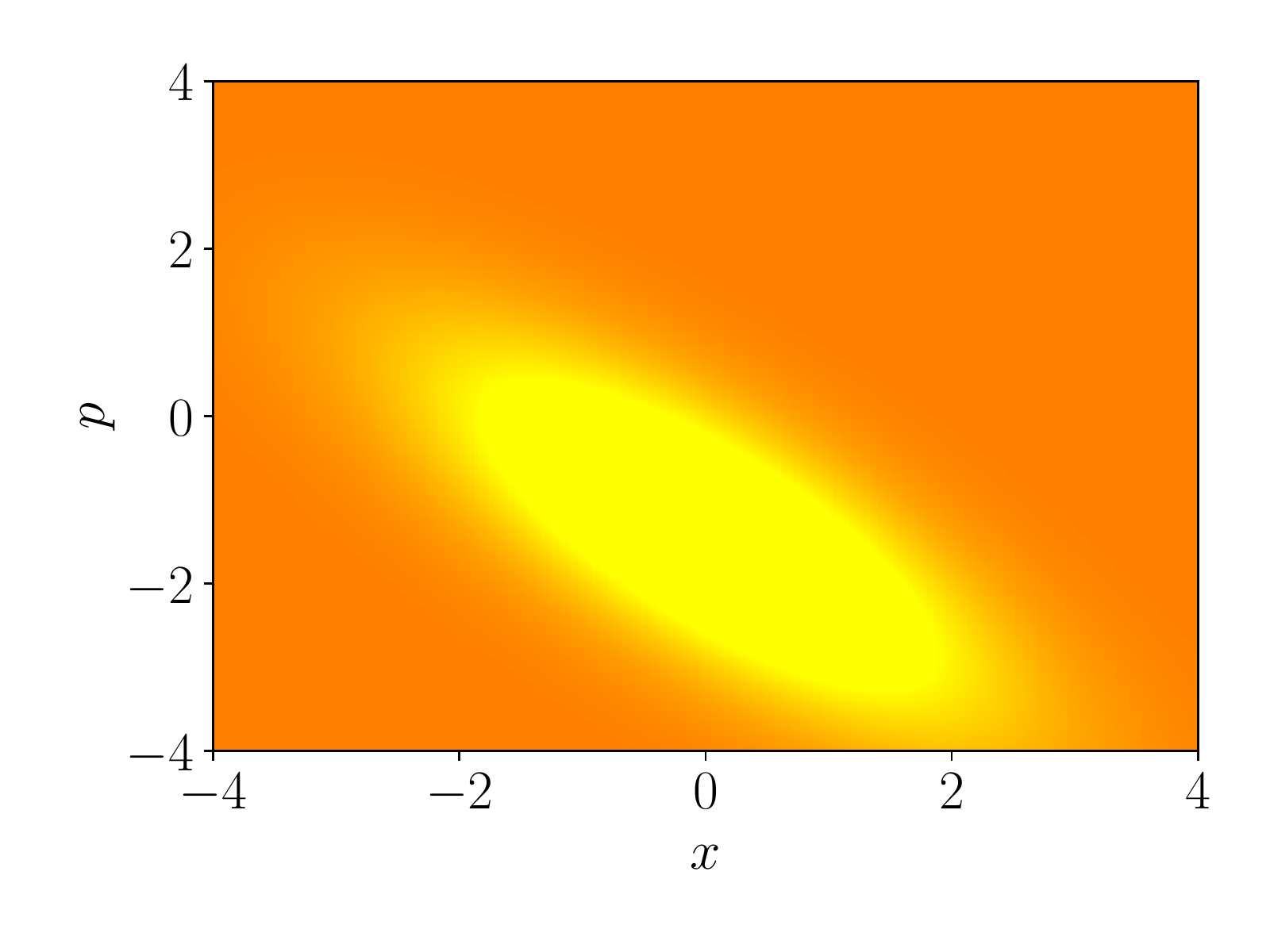} &\includegraphics[width=4.cm]{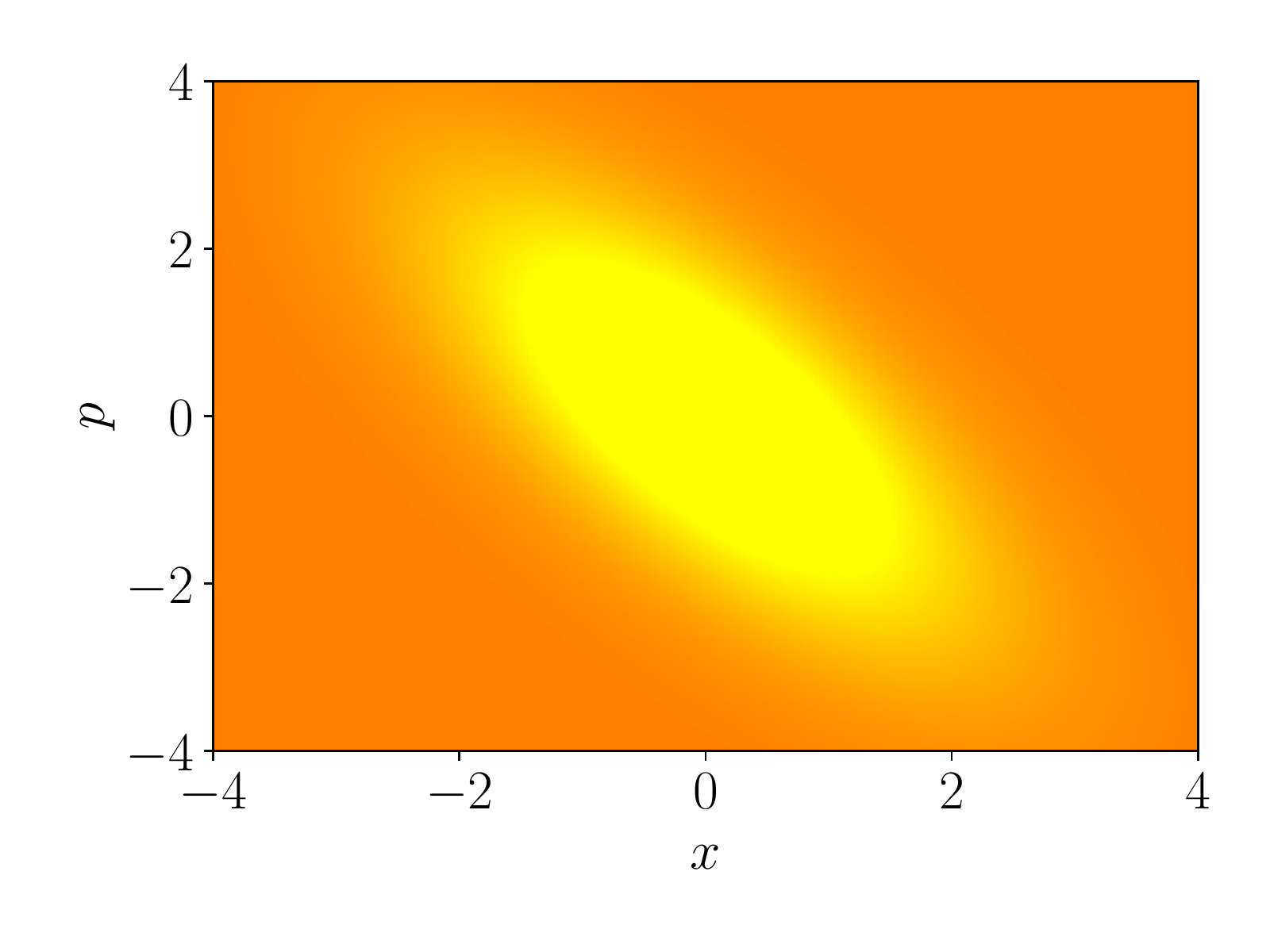} &\includegraphics[width=4.cm]{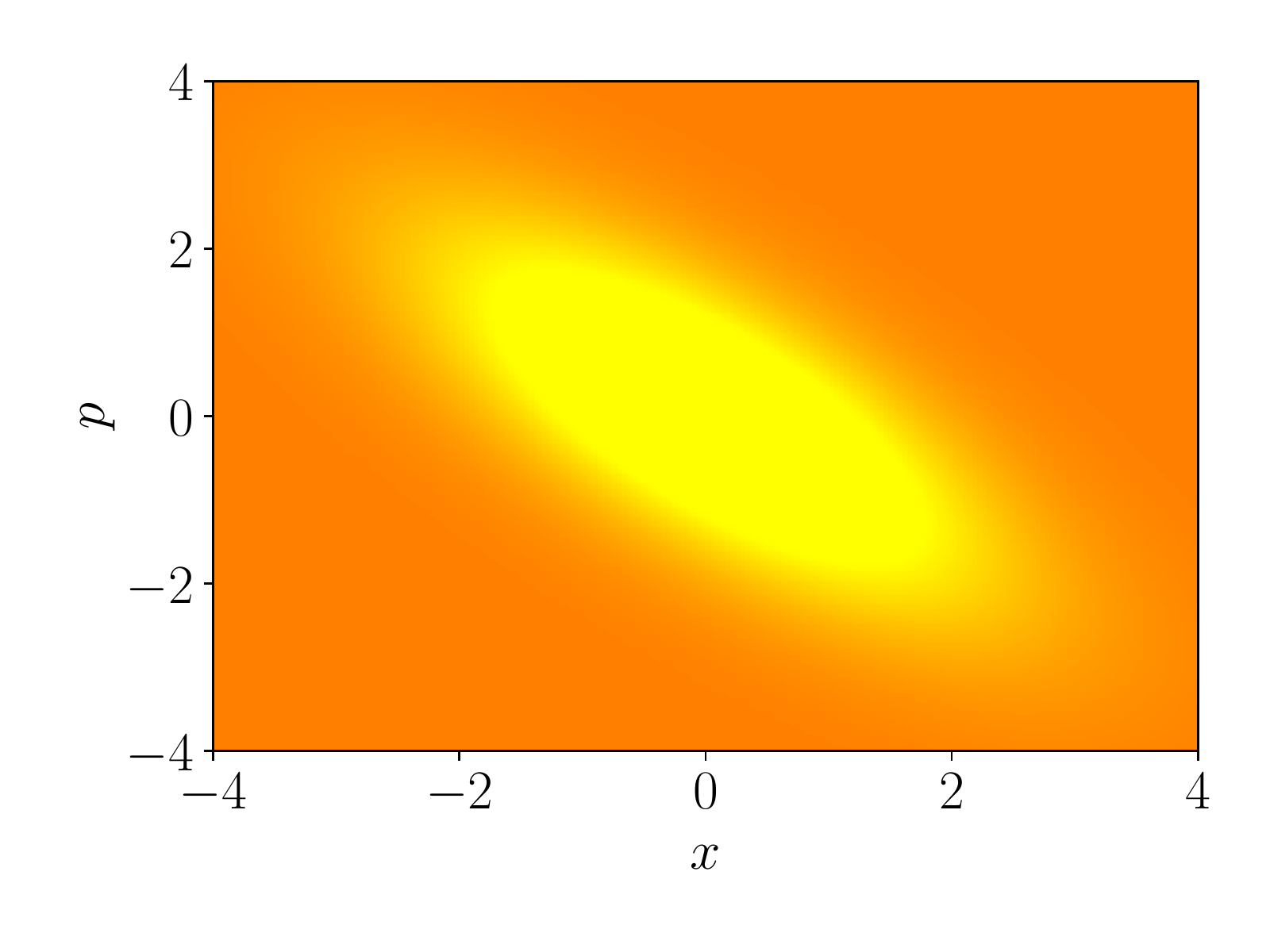} &\includegraphics[width=4.cm]{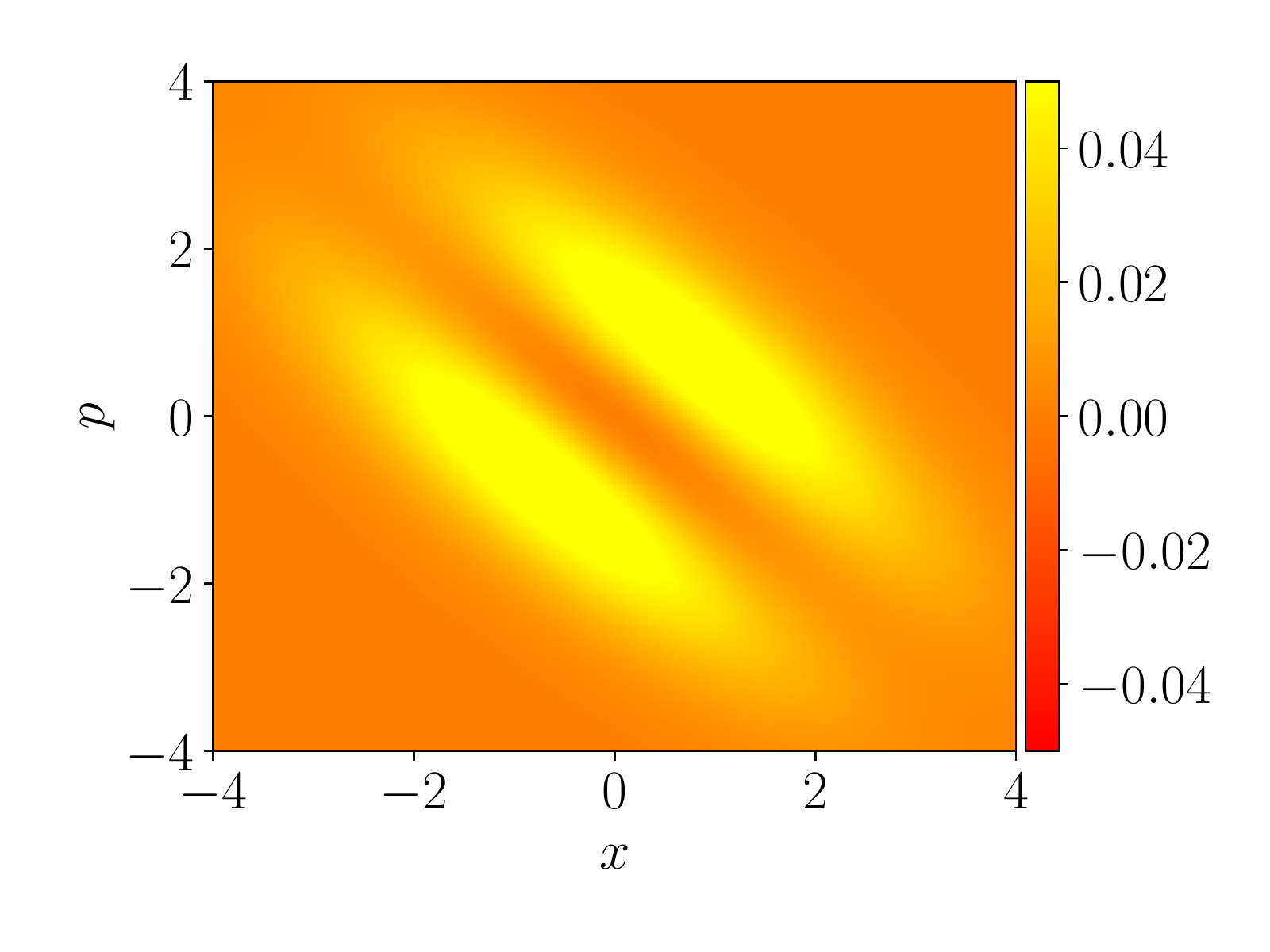}\\
		\vspace{-2.cm}Thermal state\vfill\mbox{$n_2=10^{-3}$} &\includegraphics[width=4.cm]{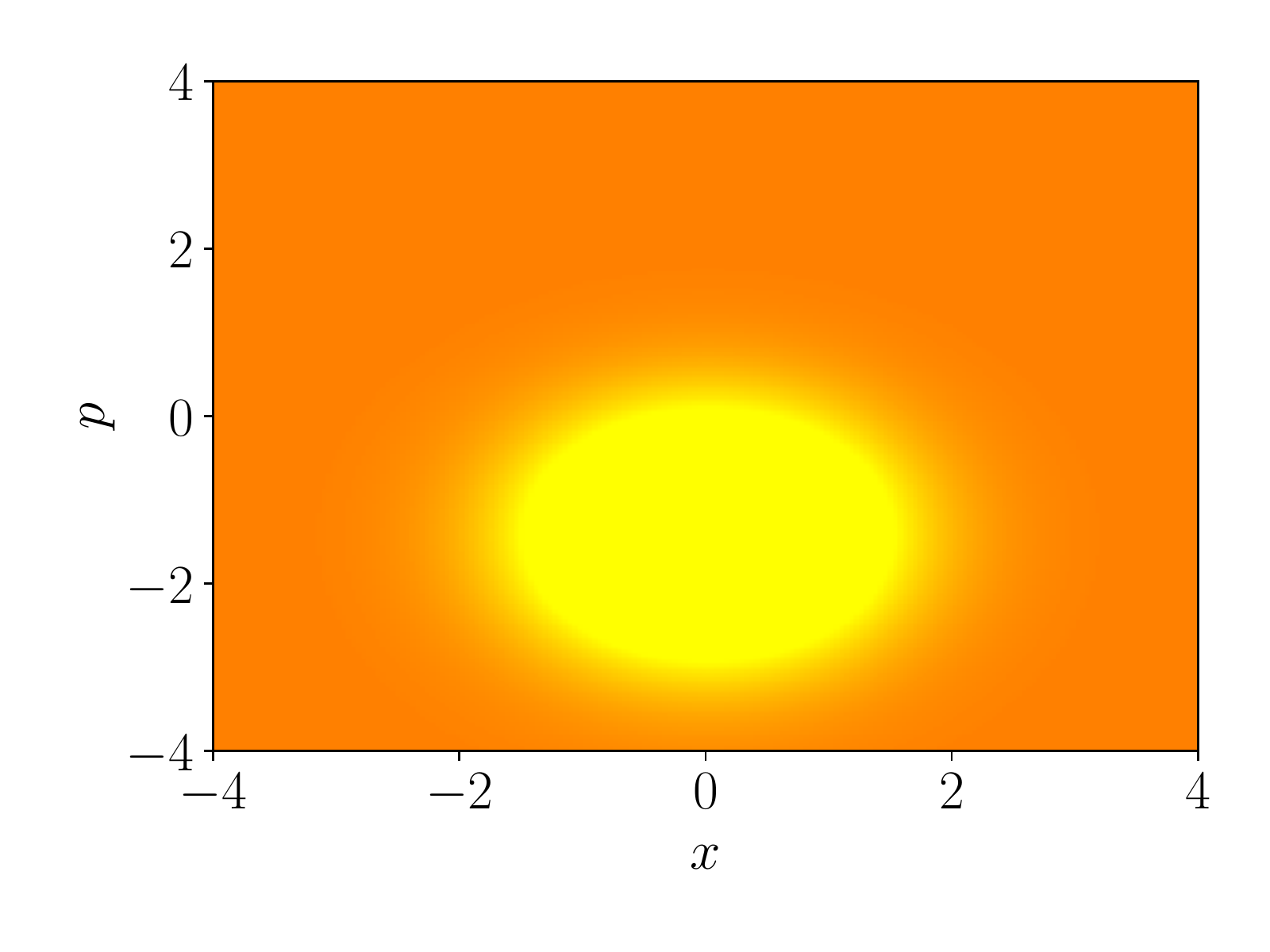} &\includegraphics[width=4.cm]{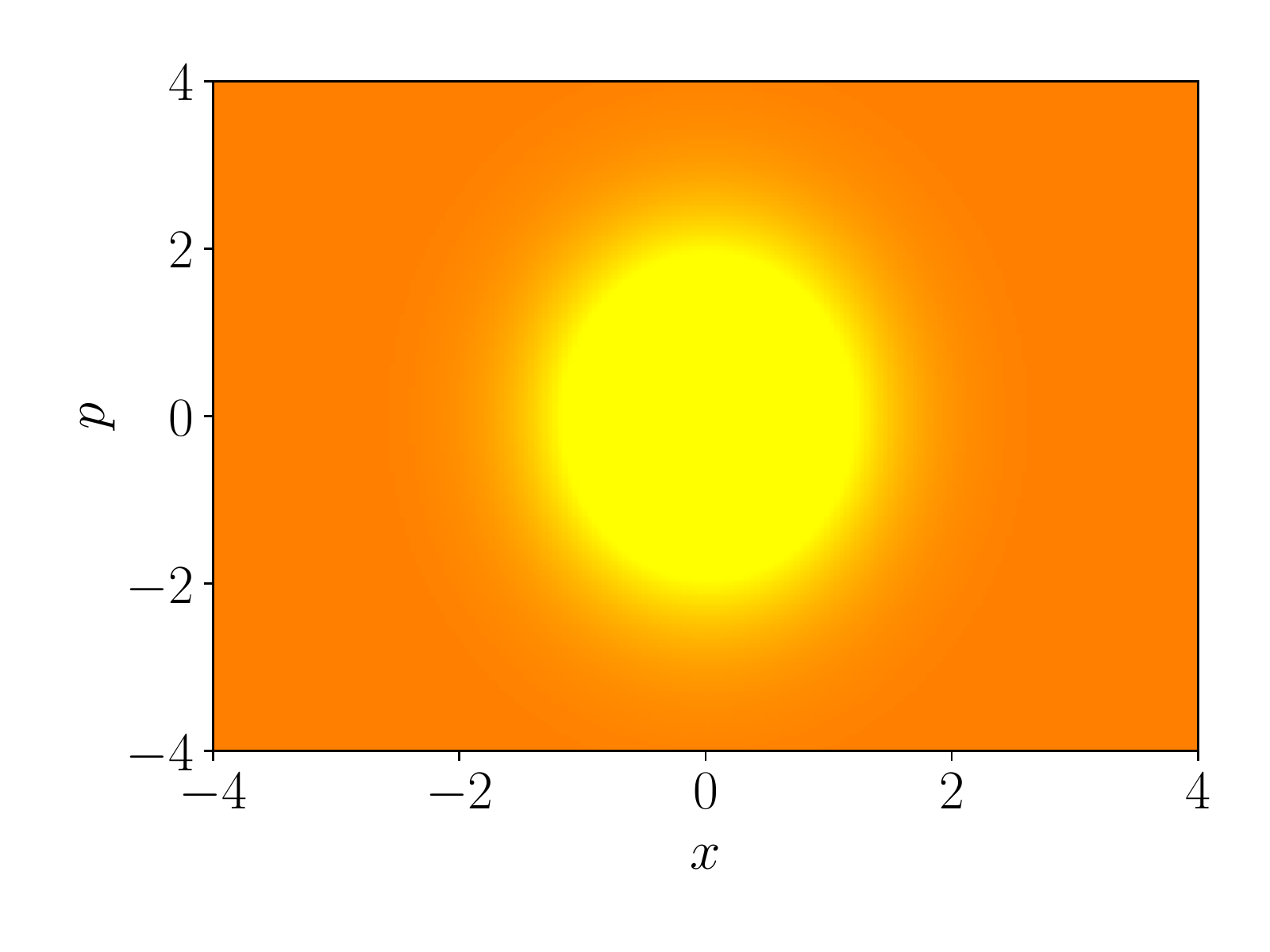} &\includegraphics[width=4.cm]{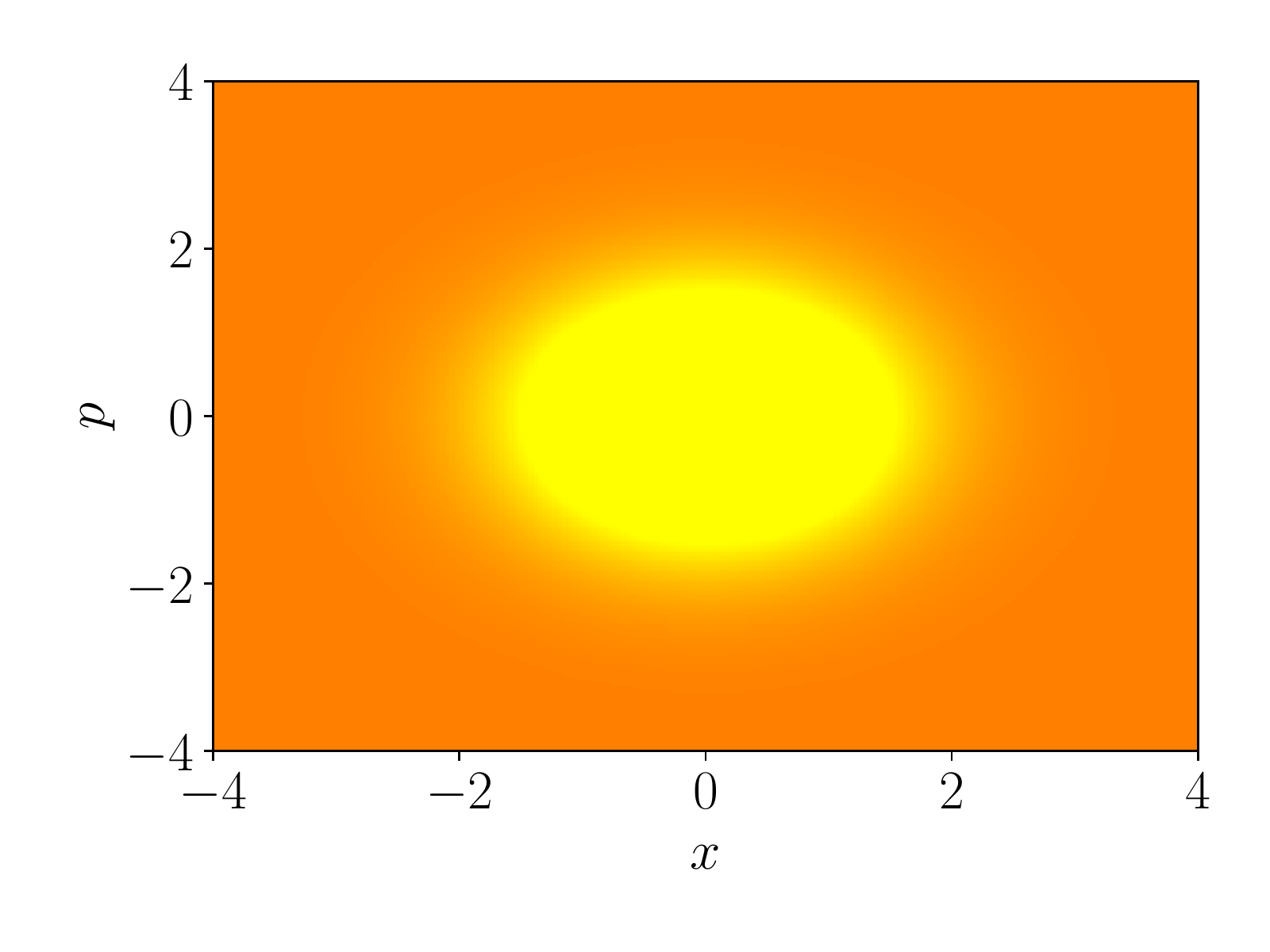} &\includegraphics[width=4.cm]{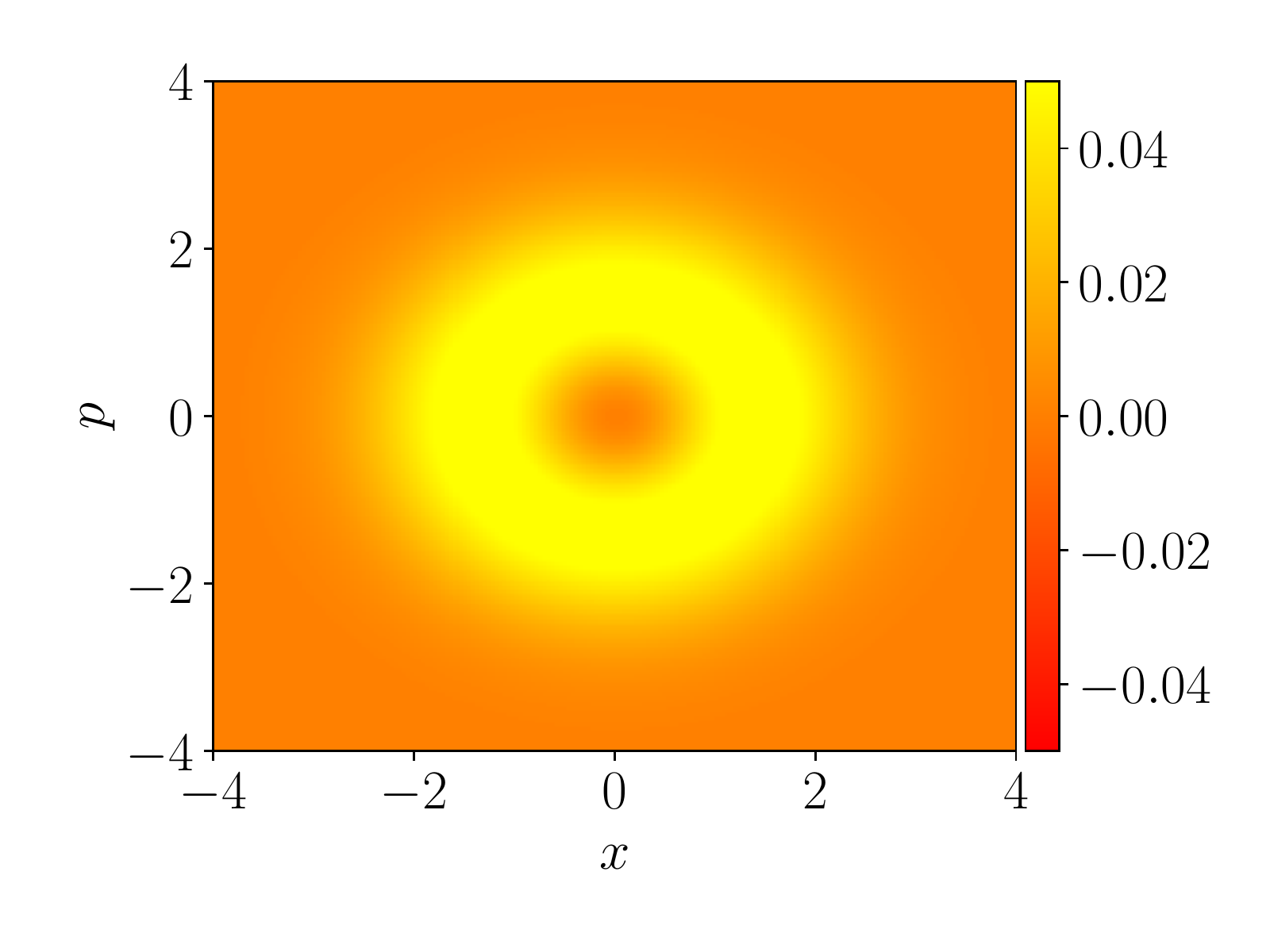}\\
		\vspace{-2.cm}Fock state\linebreak$l_1=2$
		&\includegraphics[width=4.cm]{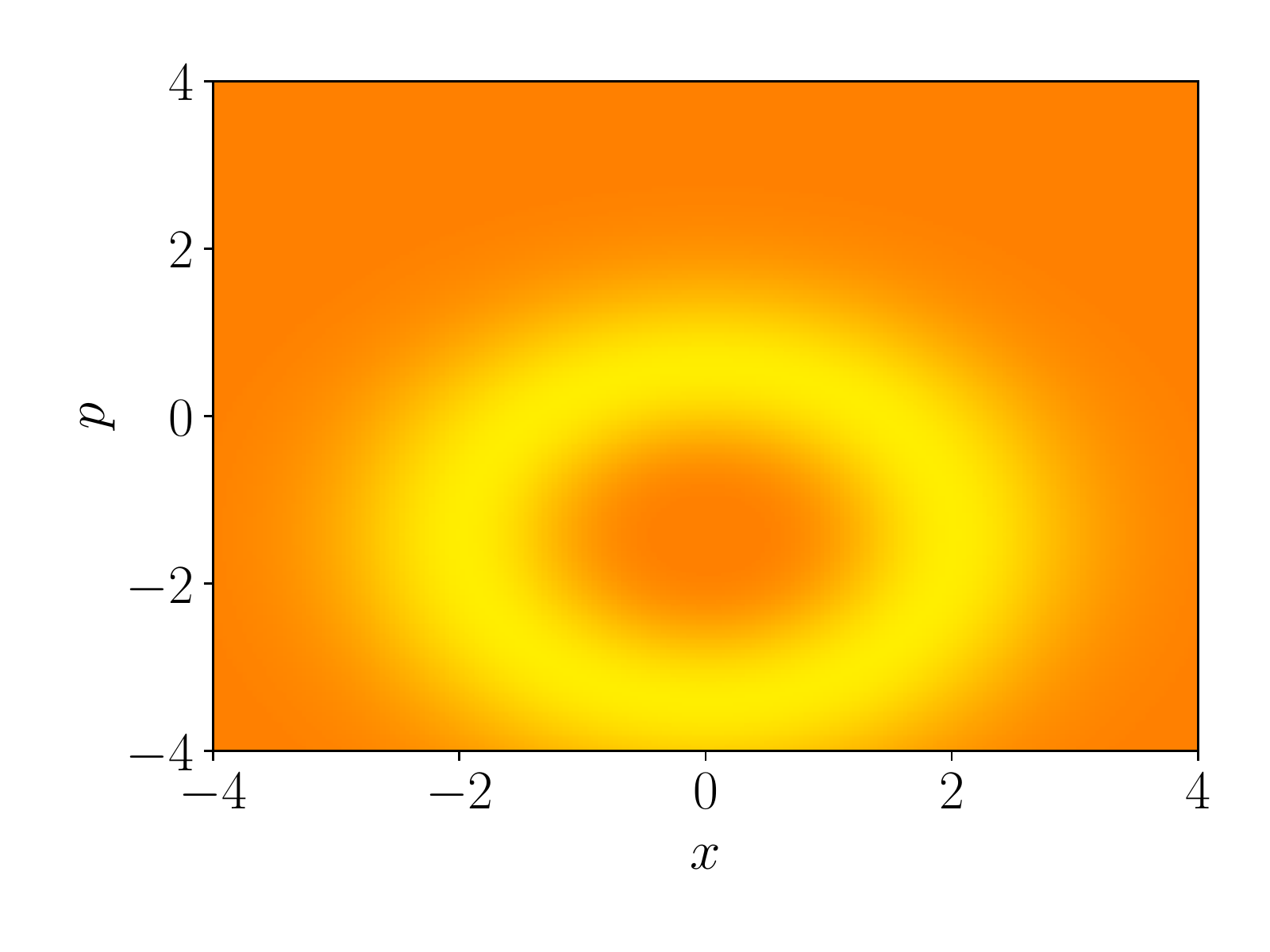} &\includegraphics[width=4.cm]{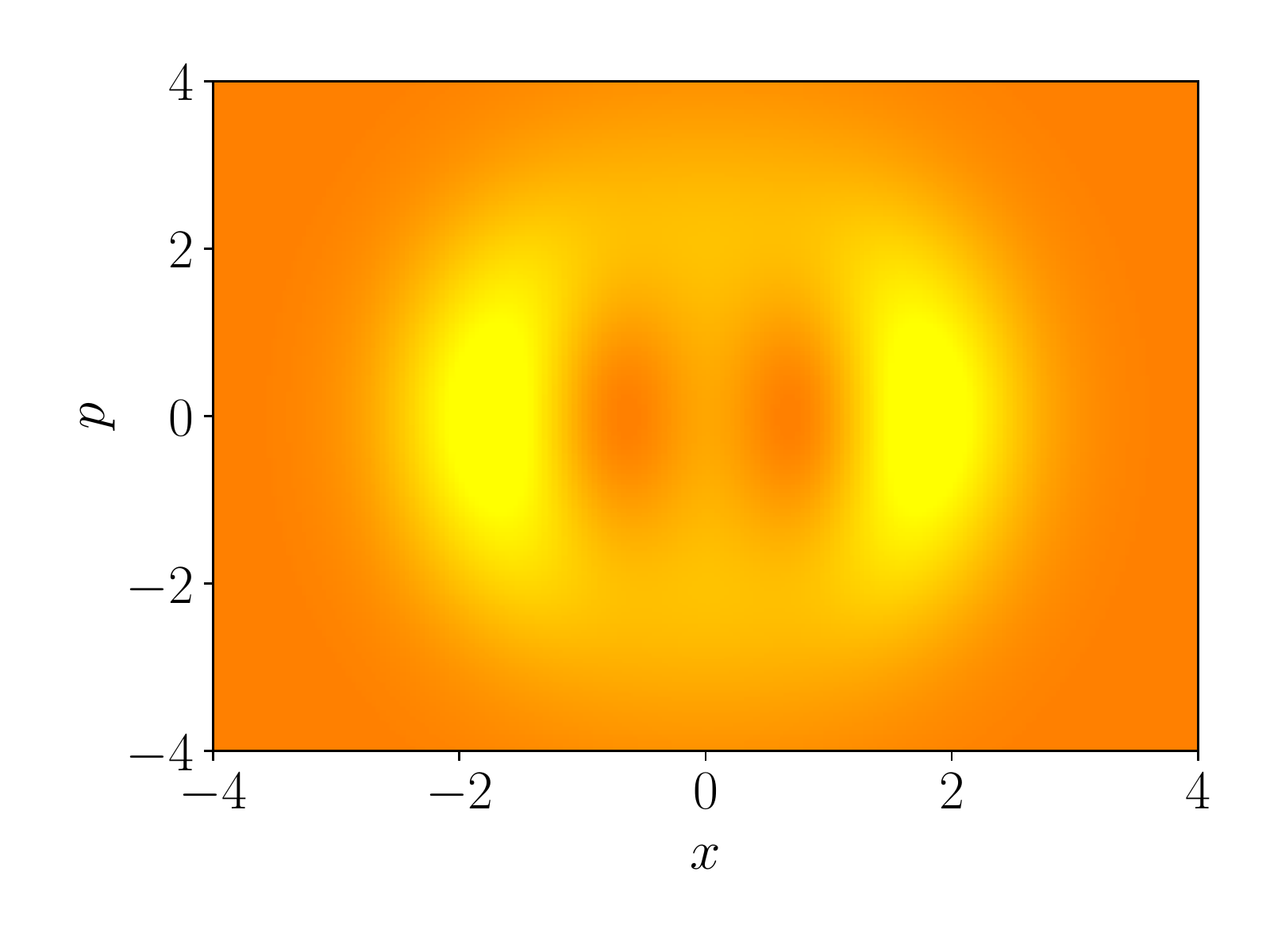} &\includegraphics[width=4.cm]{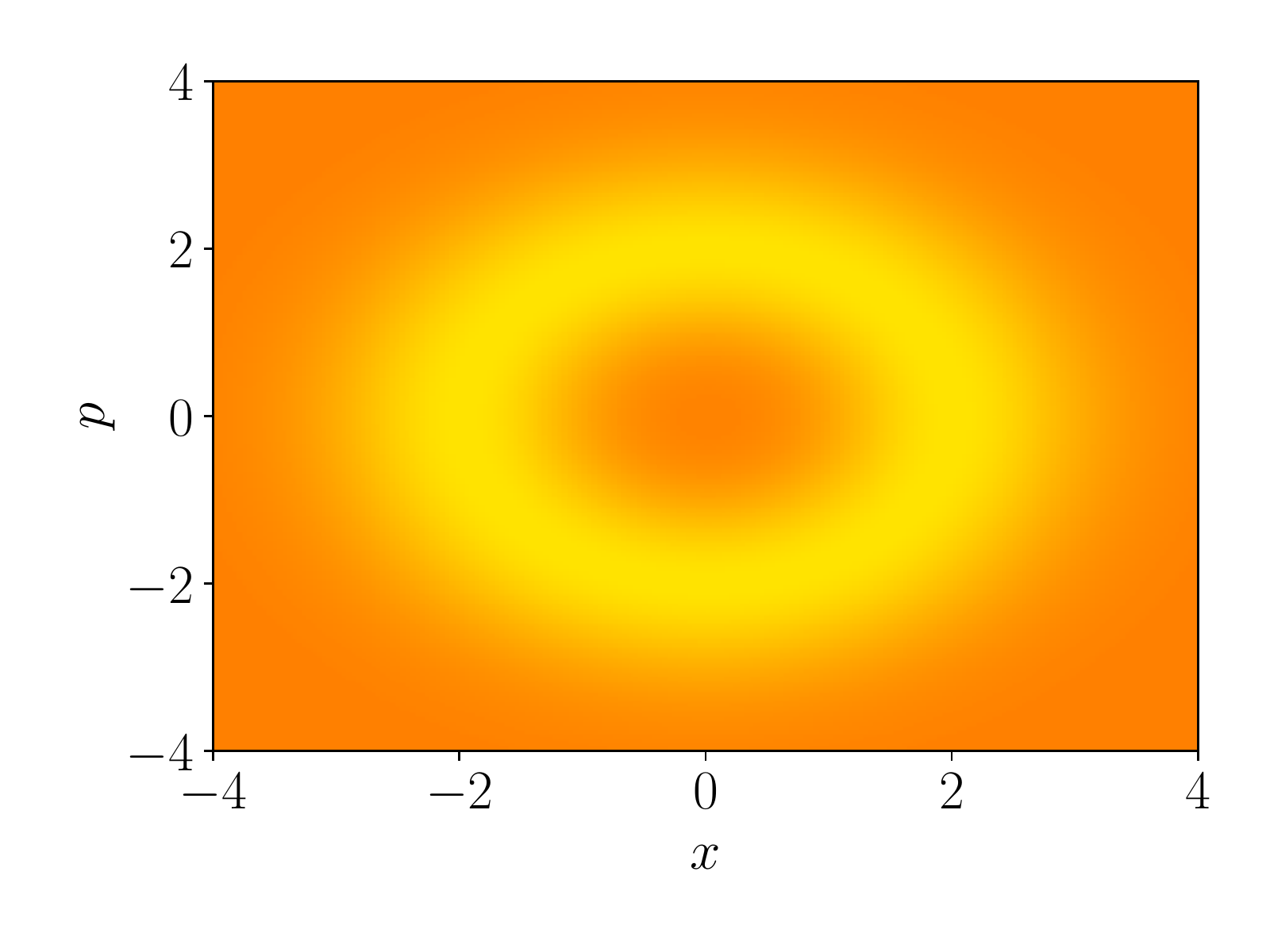}
		 &\includegraphics[width=4.cm]{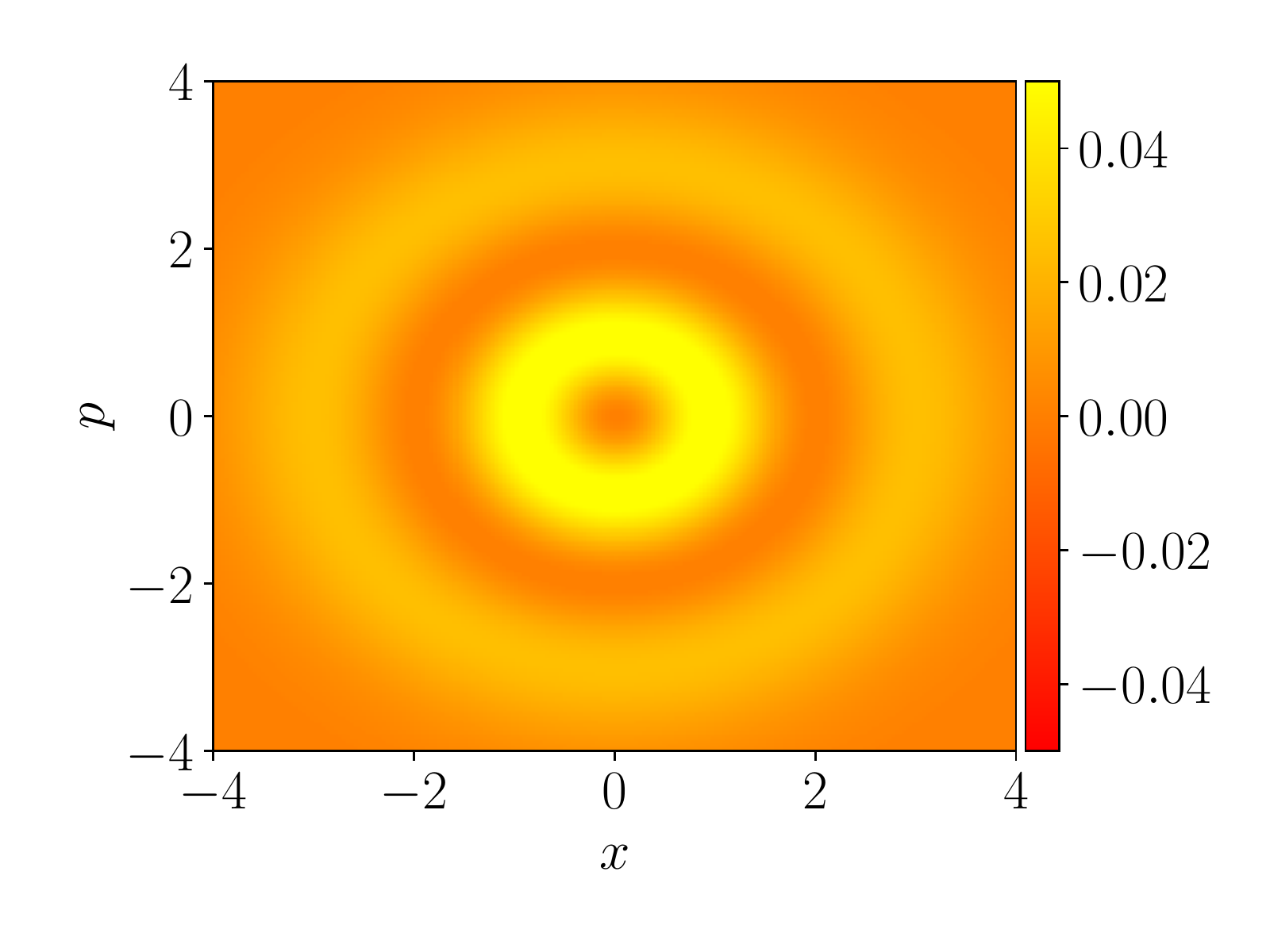}

	\end{tabular}

	\caption{Convoluted Wigner functions of the coherent, squeezed coherent, thermal and Fock state with different parameters. }

\label{tab:conv1}
	\end{table*}

\end{document}